\documentclass[showpacs,rmp,twocolumn]{revtex4}
\usepackage{graphics,graphicx,amssymb,amsmath}
\usepackage{natbib}
\newcommand{\etal}{{\it et al.}}

\begin{document}

\title{Herbertsmithite and the Search for the Quantum Spin Liquid}

\author{M. R. Norman}
\affiliation{Materials Science Division, Argonne National Laboratory, Argonne, IL 60439}

\begin{abstract}
Quantum spin liquids form a novel class of matter where, despite the existence of strong exchange interactions, spins do not order down to the lowest measured temperature.  Typically, these occur in lattices that act to frustrate the appearance of magnetism.  In two dimensions, the classic example is the kagome lattice composed of corner sharing triangles.  There are a variety of minerals whose transition metal ions form such a lattice.  Hence, a number of them have been studied, and were then subsequently synthesized in order to obtain more pristine samples.  Of particular note was the report in 2005 by Dan Nocera's group of the synthesis of herbertsmithite, composed of a lattice of copper ions sitting on a kagome lattice, which indeed does not order down to the lowest measured temperature despite the existence of a large exchange interaction of 17 meV.  Over the past decade, this material has been extensively studied, yielding a number of intriguing surprises that have in turn motivated a resurgence of interest in the theoretical study of the spin $\frac{1}{2}$ Heisenberg model on a kagome lattice.  In this colloquium article, I will review these developments, and then discuss potential future directions, both experimental and theoretical, as well as the challenge of doping these materials with the hope that this could lead to the discovery of novel topological and superconducting phases.
\end{abstract}

\date{\today}

\maketitle

\tableofcontents

\bigskip

In 1905, G. F. Herbert Smith of the British Museum of Natural History reported the mineral paratacamite, Cu$_4$(OH)$_6$Cl$_2$, a hydroxychloride
of copper \cite{smith}.  He likely did not realize the future significance this would have for the physics community.
In the years afterwards, there were a number of studies
connected with this paper, much of it driven by the desire to understand the corrosion of copper.  But in the 1980s, there
was renewed interest from the mineralogical community, given the realization that copper hydroxychlorides form a variety
of structures given the different coordinations of copper ions with (OH) and Cl ligands and its Jahn-Teller nature \cite{eby,burns}.
Then, in 2004, Braithwaite and collaborators provided a further clarification \cite{braithwaite}.  Zinc was necessary to stabilize
the rhombohedral paratacamite structure, and they proposed that the end member, ZnCu$_3$(OH)$_6$Cl$_2$, be named
herbertsmithite in honor of Herbert Smith.

For physicists, though, the story began the following year, when Dan Nocera's group at MIT synthesized crystals of this material
and then studied their magnetic properties \cite{shores}.  The motivation was that in the rhombohedral structure (Fig.~1), the copper ions form a special lattice
of corner sharing triangles known as the kagome lattice (`kagome' from the pattern one sees in Japanese basketwork).  For antiferromagnetic
superexchange interactions (which were suspected given that the Cu-O-Cu bond angles were near 120$^\circ$), this lattice is the most magnetically
frustrated in two dimensions.  This, coupled with the low spin of the copper ions (S=$\frac{1}{2}$) which maximizes quantum fluctuations as well
as the tendency towards
singlet formation, makes these materials ideal ones to search for the existence of a quantum spin liquid, a line of thought going
back to Phil Anderson's original work on this subject \cite{rvb} which came to the forefront when he suggested that a liquid of such singlets (which
he denoted as `resonating valence bonds') played a central role in the physics of cuprate superconductors \cite{rvb2}.

\begin{figure}
\includegraphics[width=0.49\hsize]{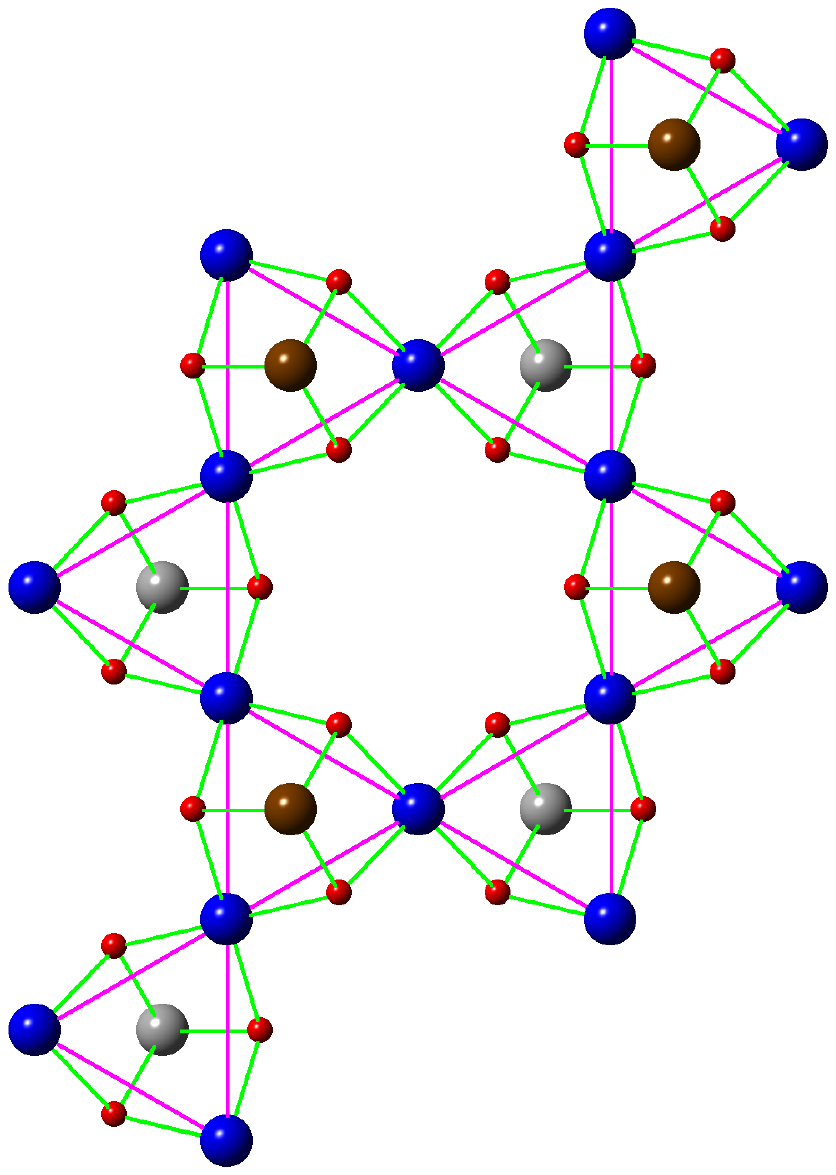}
\includegraphics[width=0.49\hsize]{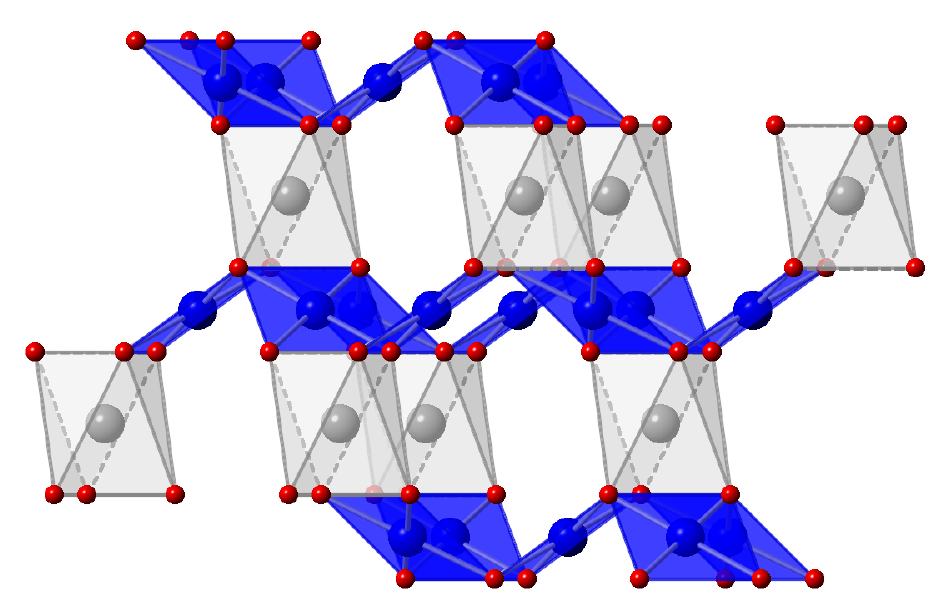}
\caption{(Color online) Left: Crystal structure of herbertsmithite, ZnCu$_3$(OH)$_6$Cl$_2$ \cite{shores}, looking down along the hexagonal $c$ axis.
Copper atoms are blue, oxygen red, and zinc atoms are gray if they sit below the copper plane, and brown if they sit above, to better
emphasize that the CuO$_4$ units form a buckled pattern (with the octahedral axis tilted 38$^\circ$ relative to the $c$ axis).
Chlorine and hydrogen atoms have been suppressed for clarity. Magenta lines emphasize the planar kagome lattice.
Right: A side view of the crystal structure using a polyhedral representation, emphasizing the ABC stacking of the kagome layers
due to coupling of CuO$_4$ planar units by ZnO$_6$ octahedra (here, all zinc atoms are shown as gray).
Note, though, that even in nominally stoichiometric compounds, there is a significant percentage of coppers sitting on the zinc sites.}
\label{fig1}
\end{figure}

But, the stoichiometric
version of the cuprates (such as La$_2$CuO$_4$), where the copper ions sit on a square lattice, are indeed N\'{e}el antiferromagnets, and
it was later realized as well that the original triangular lattice considered by Anderson in 1973 should also be long-range ordered
for the Heisenberg model \cite{huse}.  But the jury was still out in regards
to the Heisenberg kagome model, hence the profound interest in materials whose magnetic ions sit on such a lattice.

In this colloquium article, I will review the physics of herbertsmithite, and then discuss possible future directions to pursue,
noting several previous reviews on this subject \cite{mendels1,mendels2,mendels3}.
Section I provides a short introduction to frustrated magnetism and quantum spin liquids, with an emphasis on the kagome lattice,
in particular summarizing the latest numerical results and their interpretation.  In Section II, the crystal structure of herbertsmithite
and a variety of related minerals are discussed, several of which have yet to be studied in any detail.  In Section III, the focus is on 
the physical characteristics of this material, including
what is known about its ground and excited states.  In Section IV, the crucial issue of defects is discussed and their influence on
the physical properties.  Finally, in Section V, the question of chemical doping is considered, both the difficulty of doing so, and the rich
physics predicted if this were successful, along with some thoughts on where the field may be headed in regards to both experiment and theory.

\section{Introduction to Spin Liquids}

Below, I give a brief synopsis of spin liquids.  For a more extensive treatment, several excellent reviews exist
\cite{balents,moessner,misguich,chalker,savary}.

\subsection{Frustrated Magnetism}

Consider spins that have antiferromagnetic interactions \cite{mendels3}.  On the square lattice, the energy is easily minimized by the
N\'{e}el configuration of alternating up and down spins.  Although Anderson speculated that quantum fluctuations might melt the N\'{e}el lattice for spin
$\frac{1}{2}$ in two dimensions \cite{rvb2}, the overwhelming evidence, both experimental and theoretical, is that this is
not the case, though the ordered moment is significantly reduced.  For his original model of the triangular lattice \cite{rvb}, though,
the situation is more subtle.  Certainly, in the Ising case, the spins are highly frustrated, implying an extensive ground
state degeneracy.  This is easy to see since if two spins on a triangle have opposite sign, the energy for the third spin is independent
of its sign.  But in the Heisenberg case,
a compromise is possible where the spins are oriented 120$^\circ$ to one another (Fig.~2).  For an edge sharing triangular
lattice, the free energy of such a solution is comparable to that based on singlets.  On the other hand, for a corner sharing triangular
lattice, with a reduced coordination number of four,
one expects extensive ground state degeneracy, as illustrated in Fig.~2, which would act to strongly frustrate any tendency towards order.

\begin{figure}
\includegraphics[width=0.9\hsize]{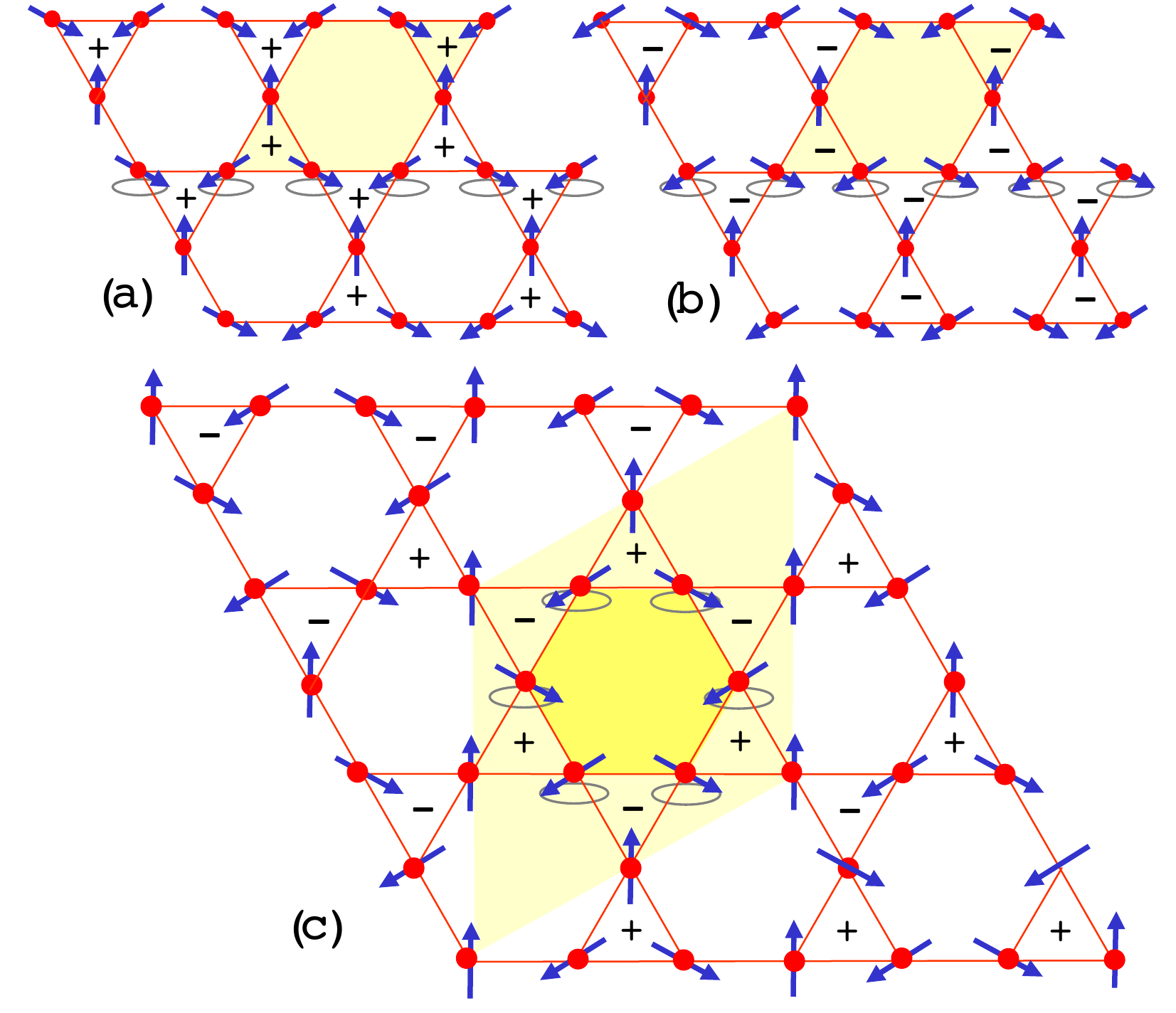}
\caption{(Color online) $q$=0 state for the near neighbor Heisenberg model on the kagome lattice for (a) positive chirality and (b) negative chirality.
(c) $q$=$\sqrt{3}\times\sqrt{3}$ state.
In each case, `zero' modes are indicated by the ellipses where the spins can turn with no cost in energy \cite{yildirim}.
}
\label{fig2}
\end{figure}

An early identifier for frustrated magnets was proposed by Art Ramirez \cite{ramirez}.  Usually, the high temperature susceptibility can
be fit to the form $\chi^{-1} \propto T - \Theta_{CW}$, where $\Theta_{CW}$ is the Curie-Weiss temperature.
In a classic antiferromagnet, one would expect the ordering temperature to be comparable
to $|\Theta_{CW}|$ ($\Theta_{CW}$ being positive for the ferromagnetic case, negative for the antiferromagnetic one).  Therefore,
the parameter $f = |\Theta_{CW}|/T_N$, where $T_N$ is the N\'{e}el temperature, quantifies how
frustrated the magnet is (more properly, how far the material deviates from mean field behavior).  For a spin liquid,
$f$ goes to infinity.  In particular, a material which should be magnetic,
but shows no ordering or spin freezing behavior down to the lowest measured temperature.

\subsection{Quantum Spin Liquids}

So, what is `quantum' about quantum spin liquids?  To understand this, we go back to Anderson's resonating valence bond (RVB) concept \cite{rvb,rvb2}.
For spin $\frac{1}{2}$, one can see that a singlet bond, with an energy -3$J$/8 (where $J$ is the superexchange
interaction), has a lower free energy than a N\'{e}el bond (-$J$/4).  Depending on the connectivity of the bonds, singlet formation can sometimes win
out.  Anderson then speculated that the free energy could be further lowered if
these singlet bonds resonated from one link of the lattice to the next (this concept being borrowed from Pauling's model for resonating carbon double bonds in a benzene
ring).  Note that a singlet involves maximally entangled spins.  Therefore, a quantum superposition of these objects would have macroscopic
quantum entanglement, implying novel topological properties \cite{balents}.  Such a resonating valence bond state would be the liquid
version of a lattice of static resonating valence bonds, known as a valence bond crystal (Fig.~3, left).  As the energetics of these two states are
similar, identifying which one is realized is one of the major topics of the field.  Experimentally, the latter can be identified as it
breaks translational symmetry, and in fact there are several candidates for this state, including the pinwheel valence bond state realized
in the kagome lattice material Rb$_2$Cu$_3$SnF$_{12}$ \cite{matan1}.

\begin{figure}
\includegraphics[width=0.49\hsize]{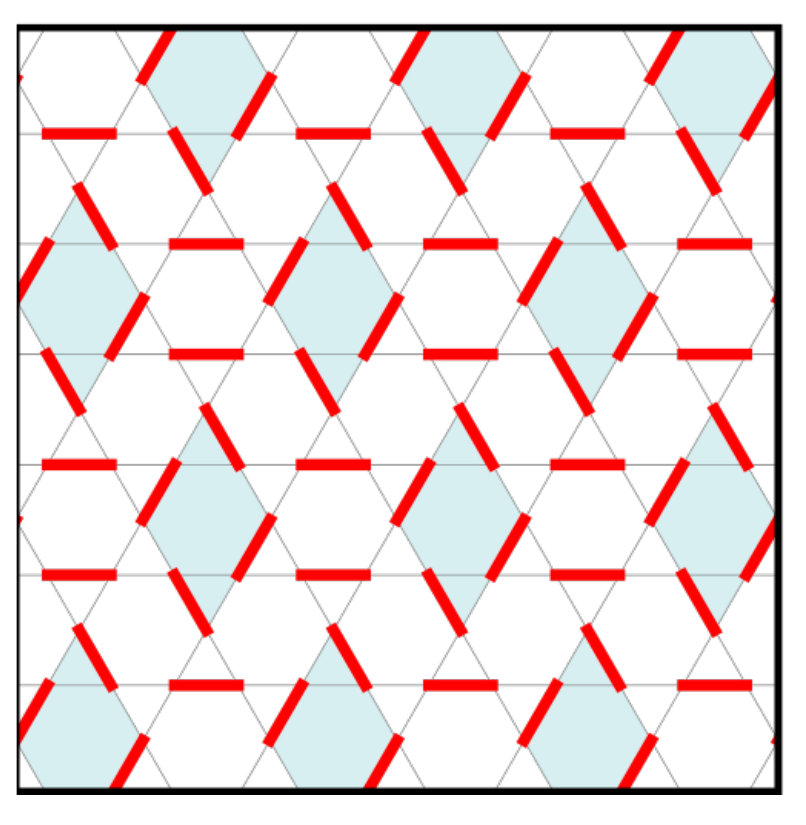}
\includegraphics[width=0.49\hsize]{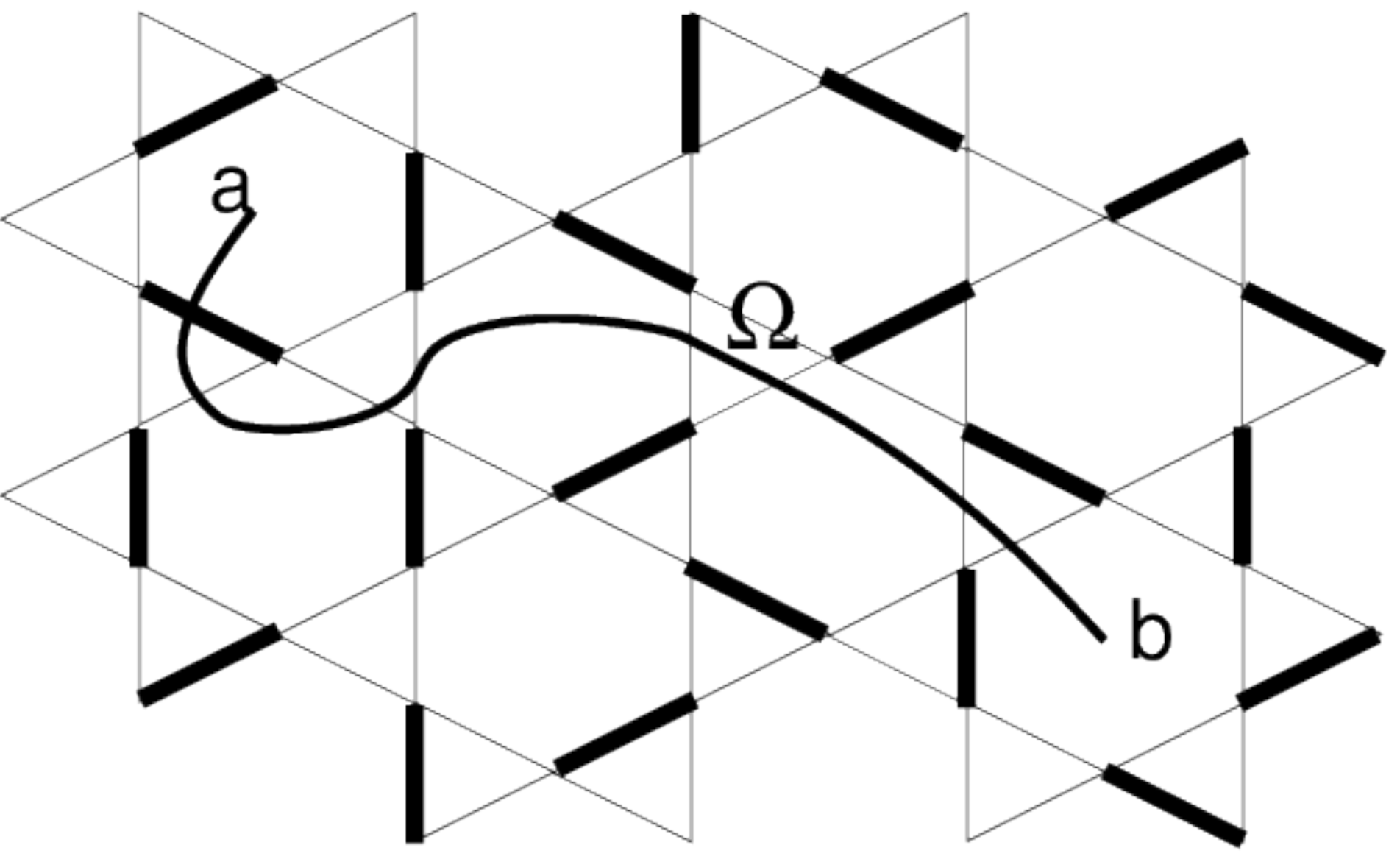}
\caption{(Color online) Left: Singlet near neighbor dimer covering of the kagome lattice.  Shown is a twelve site valence bond solid forming a diamond
pattern \cite{hwang}.  A quantum superposition of this with all other near neighbor dimer coverings would be an RVB spin liquid.
Right: A pair of visons at $a$ and $b$ \cite{misguich}.  There is a negative sign in the wave function for this excited state associated with every dimer
that the curve $\Omega$ crosses.
}
\label{fig3}
\end{figure}

Identifying a quantum spin liquid, though, is very subtle:  as emphasized in \textcite{balents}, it is easy to state what it is not rather than what it is.
Because of this, most restrict its definition to ground states with no translational symmetry breaking, but possessing non-trivial topological properties.
To date, the easiest identifier is the entanglement entropy of the system \cite{jiang}, which can probe the long-range topological properties of the ground state.  But
this is not an experimental identifier.  Rather, quantum spin liquid models are characterized by novel excitations which typically possess fractional
statistics, which can then in principle be identified by experiment as in the case of the fractional quantum Hall effect.  An easy way to see this is
to look again at Fig.~3.  Imagine that one of the sites of the lattice does not participate in a singlet bond.  This defect is known as a spinon, that
is a neutral particle with spin $\frac{1}{2}$.  Spinons can also be created in pairs by breaking apart a singlet bond.  For a short ranged RVB liquid state,
these spinons can then freely propagate by a local rearrangement of the bonds along its path.  They differ from the spin 1 magnon excitations of an ordered
magnet (the local analogue of which is exciting one of the singlets in Fig.~3 to a triplet).
Spinons are not the only potential excitations.  One can also have vortex-like
excitations known as visons \cite{vison}, where the quantum mechanical phase in the wave function associated with the dimers twists around this defect (Fig.~3, right).
A major thrust has been to attempt to find evidence for these excitations in the spin excitation spectrum measured by inelastic neutron scattering \cite{han1,punk}.
A clever experiment was also proposed to search for visons by putting magnetic flux through a loop of the material and looking for a novel type of flux quantization
\cite{senthil}.  In cuprates, this was tried with a null result \cite{moler}, but similar experiments have yet to be performed for candidate spin liquid materials.
Recently, another experiment was proposed to search for fractionalized spinon excitations by looking for coherent oscillations versus voltage in the tunneling density of states
between a superconductor or non-collinear magnet
and a spin liquid, with a result that depends on the type of spin liquid and thus the topologically non-trivial nature of the boundary the material possesses \cite{berg}.

There are many triangular-type lattices where spin liquid phases might be found.  In two dimensions, there are the honeycomb ($z$=3, where
$z$ is the coordination number), kagome ($z$=4), maple leaf ($z$=5) and triangular ($z$=6) lattices.  In three dimensions, there are the hyperkagome
and pyrochlore lattices.  Materials are known where copper ions sit on all such lattices, many of which either have suppressed
ordering temperatures or no magnetic ordering at all (or else have yet to be studied).  A few of these have been recently reviewed, including the organic
conductors where coppers sit on a distorted triangular lattice \cite{zhou}.  But for the
purposes of this article, the discussion will be restricted to that relevant for herbertsmithite and its relatives.

\subsection{The Kagome lattice and zero modes}

Returning to Fig.~2, one notes that on the kagome lattice, one can rotate spins relative to other spins without costing any energy.  This implies the
presence of zero modes, leading to an extensive ground state degeneracy.  Identifying these zero modes has some importance.
Perturbations, such as the Dzyaloshinski-Moriya (DM) interaction to be discussed below, have the effect of pushing this zero mode to finite energies where
it can be identified by inelastic neutron scattering \cite{yildirim}.  This has been observed in the iron jarosites, where the iron
ions sit on a kagome lattice \cite{jaro1}.  A sharp mode has also been seen in clinoatacamite \cite{lee,kim,wills2} and is thought to have the same origin.
But this mode is not present in herbertsmithite, rather one sees a broad continuum thought to be due to spinons \cite{han1}.

\subsection{To gap or not to gap - RVB, $Z_2$ spin liquids, and all that}

Broadly speaking, there are two types of quantum spin liquids, those which exhibit an excitation gap, and those which do not.
In each category, there is a mind boggling array of potential spin liquid
states.  For the gapped case, one generally expects the ground state to have some sort of non-extensive topological degeneracy.  One of the simplest
cases is the $Z_2$ spin liquid, connected with the visons mentioned above (where one can take the phase to be either +1 or -1 for a given dimer),
with the $Z_2$ index referring to whether a bond has a singlet (1) or not (0).

The gapless case can be more interesting.  Here, the spinon excitation gap closes.  If these zero energy states form a surface in
momentum space (analogous to the Fermi surface of a normal metal), then this is known as the uniform RVB spin liquid, first proposed in Anderson's
paper on cuprates \cite{rvb2}.  Instead,
the spinons could have a Dirac-like spectrum (with a linear $E$ versus $k$ relation, and thus a gapless momentum point), leading to a U(1) spin liquid.
This state was also proposed early on when the RVB model was being
studied in the cuprate context \cite{leeRMP}.

There are a variety of other states as well, perhaps the most relevant for the kagome case being the chiral and non-collinear states identified by
Claire Lhuillier's group \cite{messio}.  Their proposed phase diagram is shown in Fig.~4, where one sees that the near-neighbor Heisenberg model
for the kagome lattice sits at a special point where the $q$=0, $q$=$\sqrt{3} \times \sqrt{3}$, and non-collinear chiral `cuboc1' phases meet.
Since then, this phase diagram has been further refined with an eye to describing the properties of several of the materials discussed in this article \cite{bieri,iqbal,gong}.
The notion of chirality (Fig.~2) has deep meaning for triangular-type lattices, where one can define both scalar, $\vec{S}_1 \cdot (\vec{S}_2 \times \vec{S}_3)$,
and vector, $\vec{S}_1 \times \vec{S}_2 + \vec{S}_2 \times \vec{S}_3 + \vec{S}_3 \times \vec{S}_1$, chiralities.
Many chiral ground states are possible, particularly for the kagome lattice \cite{kumar},
with the role of chirality in real materials an active subject of study.

\begin{figure}
\includegraphics[width=0.6\hsize]{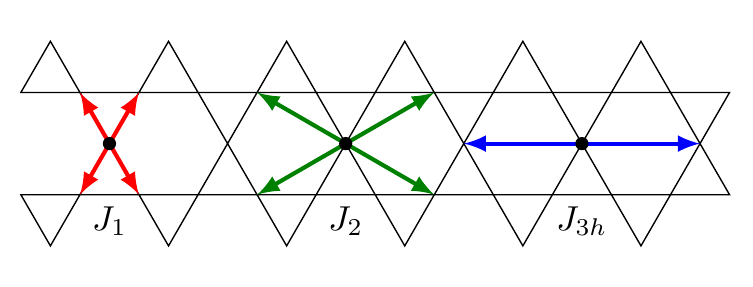}
\includegraphics[width=0.6\hsize]{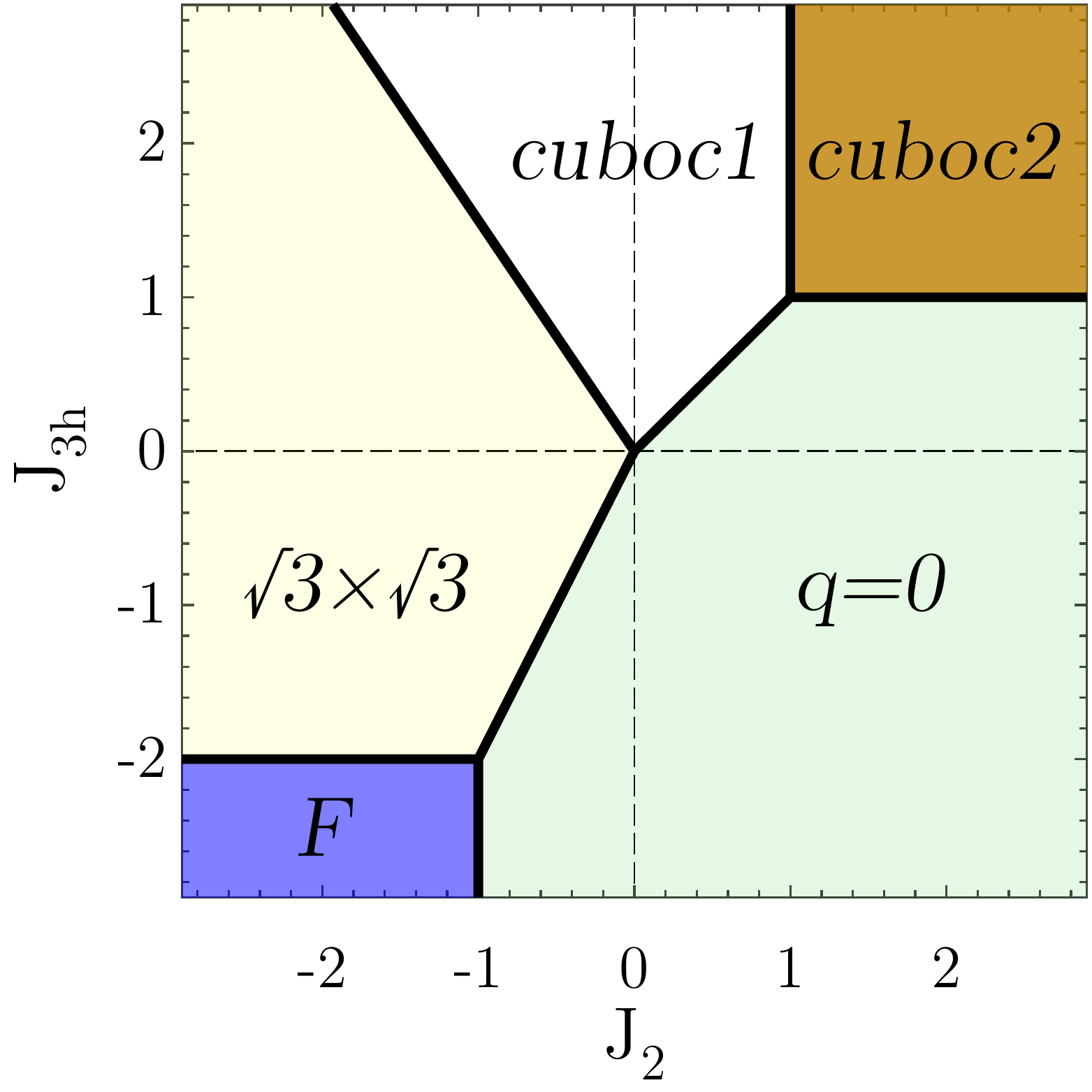}
\caption{(Color online) Phase diagram of the $J_1$-$J_2$-$J_{3h}$ Heisenberg model on the kagome lattice \cite{messio}
(here, 1 refers to near neighbors, 2 to next-near neighbors, and 3h to
coupling across a hexagon, as shown at the top).  $q$=0 and $\sqrt{3} \times \sqrt{3}$ are
coplanar spin configurations (Fig.~2), cuboc1 and cuboc2 non-coplanar ones.
}
\label{fig4}
\end{figure}

\subsection{Exact diagonalization of clusters}

Given the approximate nature of the various model states discussed above, it is desirable to have unbiased numerical evidence about the nature
of the ground state.  The first such studies involved exact diagonalization of small clusters.  Over the years, the cluster size has grown, and now
clusters up to 48 sites have been studied.  In Fig.~5, the eigenvalue spectrum for a near neighbor Heisenberg model is shown in each spin sector for a 27 site cluster \cite{lech}.
One can see
the profound difference of the triangular case from the kagome one.  This is connected with the fact that the former has long range magnetic order (and thus
a split off lowest lying `tower of states' reflecting the symmetry breaking, along with low lying magnon excitations \cite{lauchli2}), whereas the latter does not.  In particular, the kagome case has a dense array of energy levels with no obvious gap, except for the spin
gap separating the S=0 and S=1 sectors.  Extensive studies \cite{lech,sind,lauchli} have led to the conclusion that there is no gap in the singlet sector, but
a small spin gap likely exists, though the latter result has been challenged by a 42 site study \cite{nakano}.

\begin{figure}
\includegraphics[width=0.9\hsize]{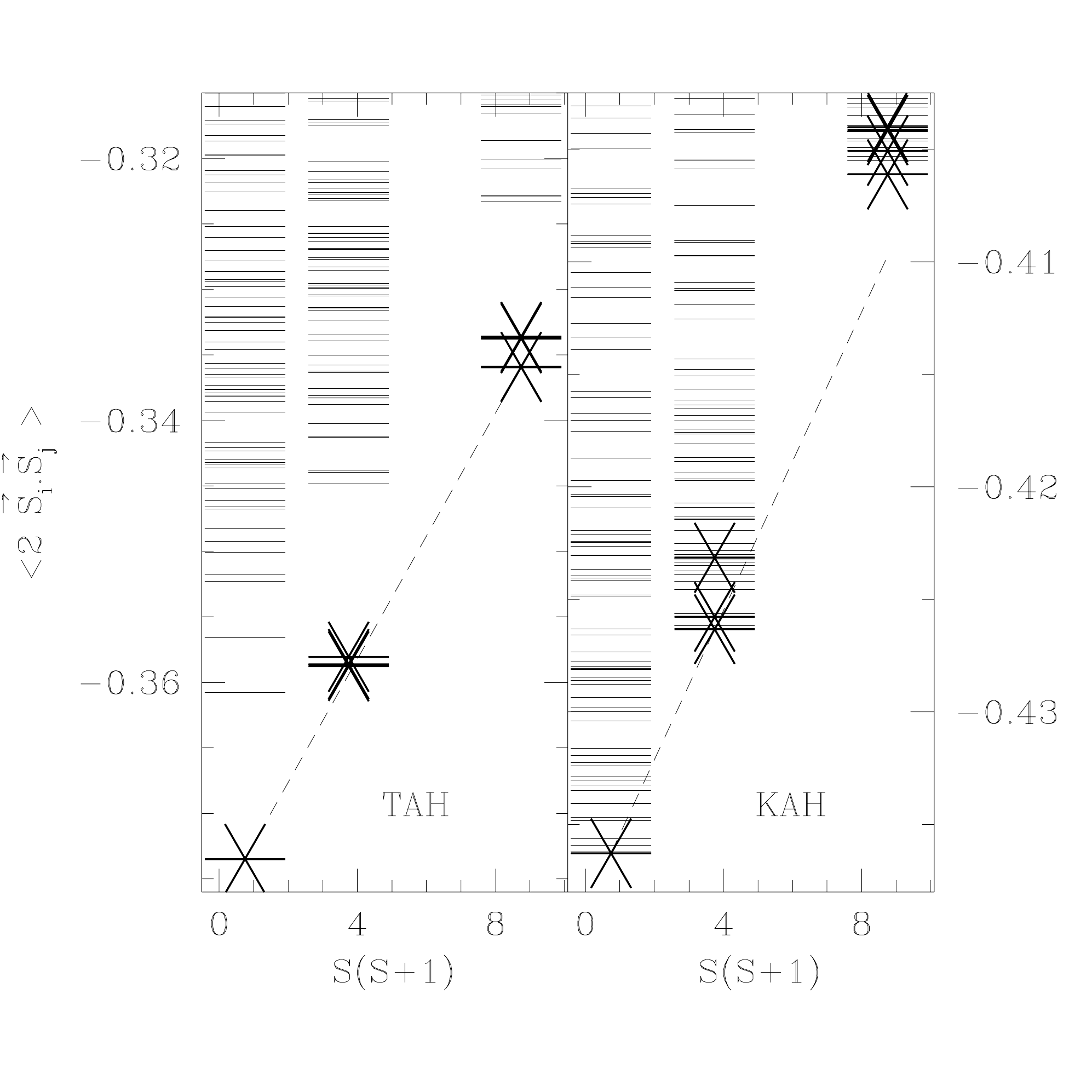}
\caption{Exact diagonalization energy spectra (27 sites) for the near neighbor AF Heisenberg model on the triangular lattice (left)
and the kagome lattice (right), as a function of the total spin of the cluster, $S$ \cite{lech}.
The crosses denote the `tower of states' that defines the ground state manifold in the infinite lattice limit.
}
\label{fig5}
\end{figure}

\subsection{DMRG, PEPS, and MERA}

To go to larger sized systems requires turning to less exact techniques.  Quantum Monte Carlo is an obvious one to suggest, but depending on the model,
one can suffer from
the famous sign problem which leads to negative probabilities in such simulations, meaning one is restricted in how low a temperature one can
study \cite{lauchli2}.  Instead, several researchers have been inspired by density matrix renormalization group (DMRG) techniques \cite{white} which have essentially led
to exact solutions for 1D spin problems.  There have been several proposals to generalize this to two dimensions.  The first is to simulate strips of the material,
which is much in the spirit of DMRG simulations in 1D.  Here, the trick is to pick the correct geometry of the strip to minimize finite size effects.  The other
two techniques, PEPS (projected entangled pair states) and MERA (multi-scale entanglement renormalization ansatz) are generalizations of DMRG to
handle extended systems in 2D, making use of much of the quantum information type approach that is the basis for DMRG \cite{SW}.  The limitation of these techniques
is that they are more biased than DMRG, and given the nearness in energy of various states for the kagome lattice, this is a potential worry.  MERA
simulations for the kagome case were reported by \textcite{evenbly} where it was concluded that the ground state was likely the 36 site valence bond crystal
proposed by Marston and Zeng \cite{marston}, and it was also noted that the free energy obtained was lower than previous DMRG studies by Donna Sheng's
group \cite{sheng}.

The situation changed in 2011 when \textcite{yan} presented extensive DMRG simulations followed by an even more detailed study from \textcite{depen}.  In these studies, the geometries of the strips were chosen so as to minimize some of the finite size effects that hampered earlier
DMRG simulations.  In particular, the studies were done on various open cylinders.  Several things emerged from these studies: (1) the spin gap was finite and (2) the ground state appeared to be a $Z_2$ spin liquid.  The latter was confirmed by studies of the entanglement entropy \cite{jiang,depen}.
Most interestingly, the response of the system was studied when strengthening certain bond patterns.  The smallest loop of resonating dimers
on the kagome lattice would be around
the six links comprising a hexagon.  The next larger loop is around the eight links comprising a diamond (Fig.~3, left).  Surprisingly, the latter had a much
larger response, and even a larger response was obtained by forming a lattice of such objects.  This implies that the $Z_2$ spin liquid identified in the DMRG
simulations is a melted version of a diamond valence bond crystal.  What should be remarked, though, is that many states have comparable free energies
(the energy difference between the $Z_2$ state and the U(1) Dirac spin liquid is only about 0.01$J$ \cite{mendels3}).

\subsection{The role of perturbations - spin anisotropy, DM interactions, and longer range exchange}

Of course, the near neighbor Heisenberg model is an idealized one.  In real systems, a host of perturbations come into play that could in principle drastically
alter the ground state.  The most obvious of these is that in most materials, the kagome network is distorted, with some of the bonds differing from others.
This typically leads to an ordered magnetic ground state, hence the search for an undistorted kagome network.  Even for a perfect network, though,
other factors come into play.  Because of spin-orbit, the spin interactions are not completely isotropic, giving rise to significant deviations from the
Heisenberg model.  In addition, in insulators, longer range exchange plays a role (Fig.~4), and in fact density functional theory (DFT) studies indicate that
next-near and even next-next-near neighbor interactions can be significant \cite{jeschke}.  And, in real materials, the 2D kagome planes are not in isolation, and again DFT studies
indicate in many cases sizable magnetic interactions between the planes \cite{jeschke}.  Finally, the most important perturbation is usually the Dzyaloshinski-Moriya (DM)
term.  For a coplanar configuration, this can act to cant the spins out of the plane.  This
perturbation is also responsible for shifting the zero mode to a finite energy \cite{yildirim}.  Finally, several studies have shown how the DM term changes
the nature of the phase diagram from that shown in Fig.~4, in some cases leading to a stabilization of the liquid phase \cite{canals,cepas}.

\subsection{Random bond models}

All the above considerations are based on perfect lattices.  But in real systems, defects play a fundamental role.  In fact, the random bond Heisenberg model
has been extensively studied in the past beginning with the work of Ma and others \cite{ma,dasgupta,bhatt}.  Upon renormalization group flow, a given
initial distribution of exchange couplings is converted to a power law distribution, leading to divergences with temperature in thermodynamic properties
such as the bulk susceptibility.  These divergences also
show up as a quantum critical like form for the dynamic spin susceptibility \cite{thill}.  Many years ago, there was a debate in the heavy fermion field whether the
observed quantum critical scaling in certain 4f and 5f electron materials
was a signature of novel Kondo physics \cite{revaz}, or due to randomness \cite{neto}.  We will have more to say about
this below when we discuss the dynamic spin susceptibility of herbertsmithite.

\section{Herbertsmithite and its Relatives}

Over the years, a variety of materials have been identified where the magnetic ions sit on a kagome lattice \cite{balents,cava,misguich,mendels3}.
Unfortunately, space precludes a discussion of these fascinating systems.
Rather, we focus here on herbertsmithite and its relatives, many of which have yet to be studied in any detail.

\subsection{Botallackite, atacamite, clinoatacamite}

The base material that leads to herbertsmitite is Cu$_4$(OH)$_6$Cl$_2$.  It exists in at least four polymorphs (see Table I), although the fourth one (a
low symmetry triclinic structure) has hardly been studied \cite{malcherek}.  The least stable (and therefore the first to form during the corrosion of
copper by sea water) is botallackite, which consists of distorted triangular copper layers with AA stacking \cite{hawthorne}.  This material exhibits long-range order at 7.2 K
\cite{zheng1}, and so will not be further discussed here.  The next most stable polymorph is atacamite \cite{parise}.
There are two different crystallographic copper sites, one associated with Cu(OH)$_4$Cl$_2$ octahedra, the other with Cu(OH)$_5$Cl octahedra.  This difference
leads to a highly distorted pyrochlore structure.  Atacamite orders at 9 K, with some evidence for spin glass behavior \cite{zheng1}. There has been one
proton NMR study challenging the purported spin glass behavior \cite{zenmyo}, but besides that, little is known about its actual magnetic structure.

The most stable polymorph is clinoatacamite (Fig.~6, left) \cite{jambor,grice}.  It has three different crystallographic copper sites.  Two of them, associated with Cu(OH)$_4$Cl$_2$
octahedra, form distorted kagome layers.  The third, associated with Cu(OH)$_6$ octahedra, sits on a distorted triangular layer and connects these
layers, leading to ABC stacking of the kagome layers, much like what is seen in the iron jarosites.  The material has interesting magnetic properties \cite{zheng2,zheng3}.  The main thermodynamic signature for ordering is at 6.5 K, and is consistent
with a transition into a canted antiferromagnetic (AF) state.  But a weak anomaly exists at 18.1 K (more on this below).  Several proposed magnetic structures exist \cite{lee,kim,wills}, with the last (shown in Fig.~6, right) most consistent with NMR data \cite{maegawa}. It is a non-collinear structure, with two of the copper sites
forming ferromagnetic chains along the monoclinic $a$-axis (AF coupled along $c$), and the other two having their spins AF oriented in the $a-c$ plane.
In addition, a net ferromagnetic component exists along the $b$-axis.  As one can see from Fig.~6, this magnetic structure is not consistent with weakly
coupled kagome layers \cite{wills}, rather the material is truly a distorted pyrochlore structure, though the former has been suggested based on a different proposed
magnetic structure \cite{helton-thesis}.  Further progress awaits single crystal studies to more definitively identify the magnetic structure.

\begin{figure}
\includegraphics[width=0.49\hsize]{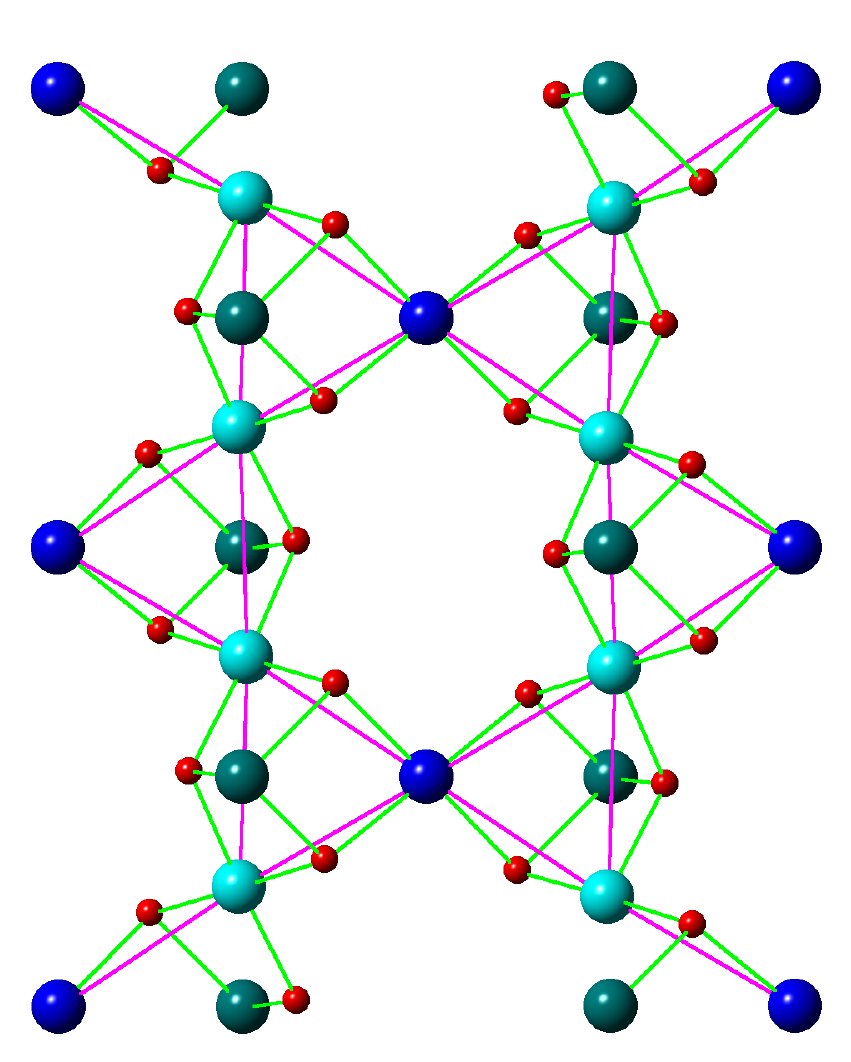}
\includegraphics[width=0.49\hsize]{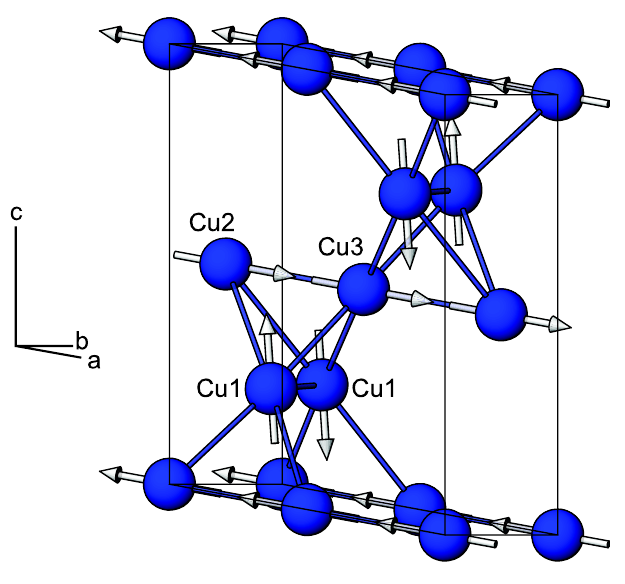}
\caption{(Color online) Left: Crystal structure of clinoatacamite, Cu$_4$(OH)$_6$Cl$_2$ \cite{grice}.  The orientation corresponds to that in Fig.~1 to emphasize similarities,
and again hydrogen and chlorine atoms have been suppressed.
There are three different crystallographic copper sites (blue, cyan, and teal), the last corresponding to the zinc site in herbertsmithite.  Note the displacement of
these `interlayer' sites relative to the center of the kagome triangles.
Right: Proposed magnetic structure of clinoatacamite \cite{wills}.  Here, Cu 1 are the cyan sites, Cu 2 the blue sites, and Cu 3 the teal sites in the left plot.
This non-coplanar magnetic structure indicates that one does not have weakly coupled kagome planes, but rather a distorted pyrochlore structure.
}
\label{fig6}
\end{figure}

Before turning to other materials, we should mention the colorfully named bobkingite, Cu$_5$(OH)$_8$Cl$_2$(H$_2$O)$_2$ \cite{bobking}.  This material
has some relation to clinoatacamite, except that double copper intersite layers connect the kagome-like sheets, with intercalated water helping to stabilize the
structure.  As a consequence of these double interlayers, the kagome-like layers have AA stacking instead.  If any magnetic studies have been done
on this material, the author is unaware of it.  One other interesting thing about this material is that the only known pure zinc analogue of the minerals
discussed here has a similar formula unit, simonkolleite, Zn$_5$(OH)$_8$Cl$_2$(H$_2$O)$_2$ \cite{simon}, though in the latter case, the space group is
the same as Zn-paratacamite.

\begin{table}
\caption{Crystallographic and magnetic properties of herbertsmithite and its relatives.  Group is the space group, and
Lat.~is the lattice on which the copper ions sit (T for triangular, P for pyrochlore, K for kagome, with K* denoting a highly distorted kagome lattice).
Order is the type of ordering (or correlations) and at what temperature (- means no order).
Note many marked AF (antiferromagnetic) also have a small F (ferromagnetic) component due to canting of the spins.
For bobkingite, $W$ stands for a water molecule.
See the text for references to the information tabulated here.}
\begin{ruledtabular}
\begin{tabular}{lllll}
Name & Formula & Group & Lat. & Order \\
\colrule
Botallackite & Cu$_4$(OH)$_6$Cl$_2$ & P2$_1$/m & T & AF (7.2K) \\
Atacamite	& Cu$_4$(OH)$_6$Cl$_2$ & Pnma & P & AF (9K) \\
Clinoatacamite	& Cu$_4$(OH)$_6$Cl$_2$ & P2$_1$/n & P & AF (6.5K) \\
Claringbullite & Cu$_4$(OH)$_6$ClF & P6$_3$/mmc & P & AF (17K) \\
Barlowite  & Cu$_4$(OH)$_6$BrF & P6$_3$/mmc & P & AF (15K) \\
Bobkingite	 & Cu$_5$(OH)$_8$Cl$_2$$W_2$ & C2/m & P & ? \\
Herbertsmithite	& ZnCu$_3$(OH)$_6$Cl$_2$ & R$\bar{3}$m & K & AF (-) \\
Tondiite & MgCu$_3$(OH)$_6$Cl$_2$ & R$\bar{3}$m & K & AF (-) \\
Kapellasite & ZnCu$_3$(OH)$_6$Cl$_2$	& P$\bar{3}$m1 & K & F (-) \\
Haydeeite & MgCu$_3$(OH)$_6$Cl$_2$	& P$\bar{3}$m1 & K & F (4.2K) \\
Zn-brochantite & ZnCu$_3$(OH)$_6$SO$_4$	& P2$_1$/a & K* & AF (-) \\
\end{tabular}
\end{ruledtabular}
\end{table} 

\subsection{Claringbullite, barlowite}

There is a related polymorph to clinoatacamite known as claringbullite (Fig.~7, left) \cite{claring}.   For a long time the actual formula unit was not understood, until
it was realized that fluorine must be present \cite{nytko-thesis}.  The accepted formula unit is now known to be Cu$_4$(OH)$_6$ClF.  A related material with
the same crystal structure, barlowite, Cu$_4$(OH)$_6$FBr, was recently studied as well \cite{barlow,barlow2}.  The biggest difference in
this structure is that instead of the copper intersites being in octahedral coordination, they instead have trigonal prismatic coordination (well known from transition
metal dichalcogenides like NbSe$_2$, but unusual for copper).  Such a structure leads to perfect kagome layers which exhibit AA stacking.  The copper intersite position, though, is further off-center than in the case of clinoatacamite, allowing for a larger Jahn-Teller effect (in this case, there are three possible
intersite locations due to the P6$_3$/mmc symmetry, each of which is one-third occupied, meaning the intersites are likely highly disordered).  In
claringbullite, one finds magnetic order at 17 K \cite{nytko-thesis}, with a slightly lower temperature of 15 K in barlowite \cite{barlow}.    Since the
possibility of claringbullite-like stacking faults has been suggested based on Raman data in Zn-paratacamite \cite{sciberras-thesis}, this leads to the
speculation that the weak magnetic anomaly seen at 18 K in clinoatacamite could be due to such stacking faults.

\begin{figure}
\includegraphics[width=0.49\hsize]{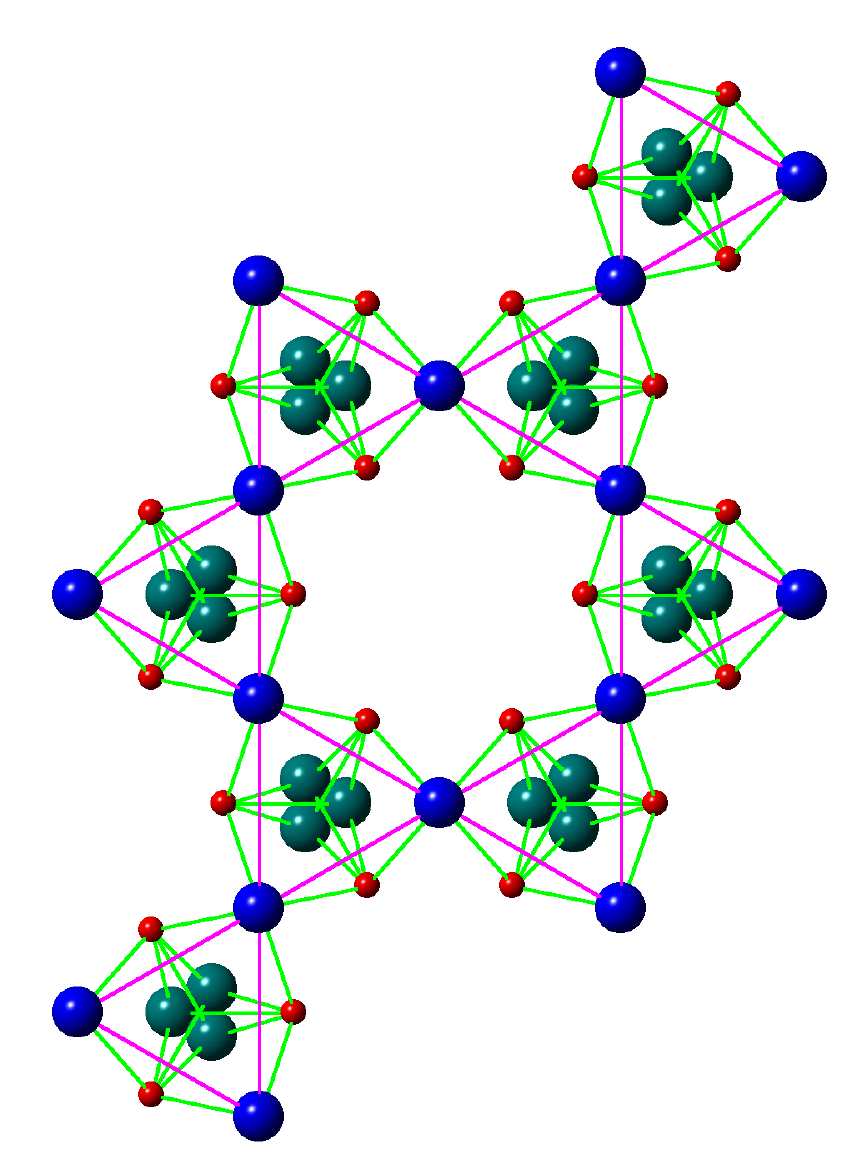}
\includegraphics[width=0.49\hsize]{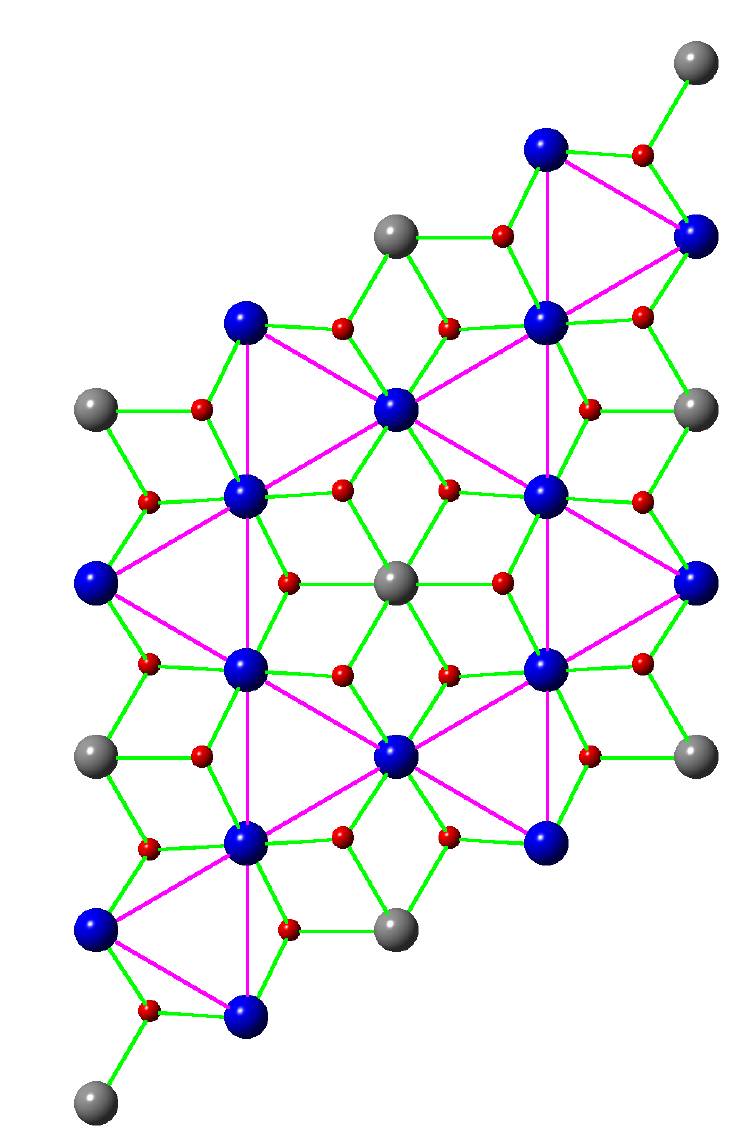}
\caption{(Color online) Crystal structures of claringbullite \cite{claring}, Cu$_4$(OH)$_6$ClF (left), and kapellasite \cite{colman2}, ZnCu$_3$(OH)$_6$Cl$_2$ (right).
The orientation is as in Fig.~1, and again hydrogen, chlorine and fluorine atoms have been suppressed.
For claringbullite, note the trigonal prismatic coordination of the copper intersites (teal) which are disordered over three
crystallographically equivalent locations.  For kapellasite, note that the zinc atoms (gray) sit in the copper planes at the middle of the hexagonal hole of
the kagome lattice.
}
\label{fig7}
\end{figure}

Attempts to `dope' zinc into the claringbullite structure has failed.  Forcing zinc in leads to a conversion to the paratacamite structure
instead \cite{shores,nytko-thesis}.  This is likely linked to the fact that zinc is not a Jahn-Teller ion, and so prefers octahedral as compared to trigonal
prismatic coordination.  Similar attempts have not been reported yet for barlowite.

\subsection{Zn-paratacamite, herbertsmithite, tondiite}

Zinc, though not readily uptaken, can be forced into clinoatacamite, leading to Zn$_x$Cu$_{4-x}$(OH)$_6$Cl$_2$.  Once $x$ exceeds about 1/3,
the monoclinic P2$_1$/n clinoatacamite structure converts to the rhombohedral R$\bar{3}$m structure with undistorted kagome layers (Fig.~1) \cite{braithwaite,shores,mendels1}.
To within experimental accuracy, zinc goes solely onto the copper intersite positions \cite{freedman}.  Close to the 1/3 crossover, an R$\bar{3}$ superstructure
has been identified \cite{fleet}, and in fact one can go reversibly from one structure to the other \cite{welch}.  This superstructure is characterized
by two different intersites, one octahedral, the other distorted much like in clinoatacamite, though to date, there is little evidence that copper goes on one
site and zinc on the other.  Away from the 1/3 crossover regime, there is no evidence for the superstructure anymore \cite{nytko-thesis}, but presumably
there are still local distortions about the copper intersites, as indirectly inferred from $^{35}$Cl NMR data \cite{imai1,fu-thesis}.

There is some controversy in the field whether a true ZnCu$_3$(OH)$_6$Cl$_2$ phase actually exists.  Resonant x-ray data are consistent with about 15\%
copper on the zinc sites with no zinc on the copper kagome sites, implying an actual formula unit of Zn$_{0.85}$Cu$_{3.15}$(OH)$_6$Cl$_2$ \cite{freedman}.
On the other hand, later synthesis studies have claimed to be able to go all the way out to $x$ = 1.16 \cite{devries},
implying a complete filling of zinc on the intersites with a significant number of zinc ions sitting on the copper kagome sites.  Certainly, the large single crystals
that have recently become available \cite{chu1,han2} appear to be zinc deficient but with pristine copper kagome planes.

Other ions besides zinc can be put on the intersites.  The most common one is magnesium, first called Mg-herbertsmithite \cite{chu2,colman1,colman-thesis}, but
now known as tondiite.  Similar issues exist with magnesium stoichiometry on the intersites.  In addition, nickel and cobalt can go on the intersites, but as
these ions are magnetic, we will not discuss them here.  Cadmium can also be accommodated, but because of its larger ionic radius, the
crystal structure distorts in such a way that the kagome-like layers are no longer oriented perpendicular to the hexagonal $c$-axis \cite{mcqueen}.

\subsection{Kapellasite, haydeeite, centennalite, Zn-brochantite}

There is a metastable polymorph to herbertsmithite known as kapellasite \cite{malcherek2, colman2, colman-thesis}.  The structure, though, is different,
with no intersites (Fig.~7, right).  Rather, the zinc ions sit in the hexagonal hole of the kagome layers (the kagome layers have AA stacking
as well).  As a consequence, the Cu-O-Cu bond angles
are smaller than in herbertsmithite, leading to ferromagnetic near neighbor correlations.  But the zinc ions also intermediate a
longer range magnetic interaction within the kagome layers which leads to frustration.
The net result is that kapellasite has no long range order.  A magnesium version (haydeeite) is also known which does order
ferromagnetically at 4.2K \cite{colman3,colman-thesis}.  A calcium version, centennallite, is also known \cite{sun}.
Kapellasite has been studied by inelastic neutron scattering \cite{fak}, and is proposed to have short range correlations related to the
non-collinear `cuboc2' structure introduced in \textcite{messio}.  This is consistent with a recent DMRG study as well \cite{gong}.
Haydeeite has also been studied by neutrons \cite{boldrin}, and the conclusion is
that although it is an ordered ferromagnet, it is also close to the phase boundary with the `cuboc2' state.

A related material has copper kagome layers connected by organic linkers, Cu(1,3-bdc) \cite{nytko,nytko-thesis}.  This material orders at 2 K \cite{nytko}
and has been measured by neutron scattering \cite{chisnell},  which reveals ferromagnetic kagome layers coupled antiferromagnetically along
the $c$-axis
(the material exhibits AA stacking).  Its spin waves have also been mapped out by inelastic neutron scattering \cite{chisnell}.  Interestingly,
a large thermal Hall effect has been seen in this material \cite{hirsch}, and both these experiments have been interpreted in
terms of novel topological properties associated with a ferromagnetic kagome lattice.

Finally, magnetic studies have been done on Zn-brochantite, ZnCu$_3$(OH)$_6$SO$_4$ \cite{li}.  This material is somewhat reminiscent of kapellasite,
with the Zn ions sitting in the middle of the kagome hexagons.  But the kagome layers are both highly distorted and strongly buckled (the space group
is P2$_1$/a), these layers being interconnected by SO$_4$ tetrahedra.
This material has some similarities with herbertsmithite, including a significant percentage of coppers sitting on the zinc sites, and a dynamic
spin susceptibility from neutrons exhibiting quantum critical-like scaling, at least at higher temperatures \cite{zorko2}.

From here on, we will focus on herbertsmithite, as it has been by far the most studied of these materials.

\section{Physical characteristics of Zn-paratacamite and herbertsmithite}

\subsection{Magnetic properties}

The original work of \textcite{shores} found that the Curie-Weiss temperature increased in magnitude as one moves from clinoatacamite towards
herbertsmithite, with a value of -314 K for the latter.  Despite this, no sign of magnetic order for the latter was found down to less than 2K
(see also \cite{bert}).  Subsequent
$\mu$SR measurements down to 20 mK also found no order for either herbertsmithite or for Zn-paratacamite ($x$=0.66)  \cite{mendels0}.  INS measurements
for herbertsmithite at 35 mK \cite{helton1} found a divergent susceptibility which mirrored the bulk susceptibility, but again no order.  Interestingly, the specific heat \cite{helton1,devries1}
indicated gapless behavior, but with a strong field dependence which is now understood to be due to copper defects on the zinc sites (Fig.~8, left).
The $^{35}$Cl Knight shift \cite{imai1}
did not follow the bulk susceptibility (though its line width did), rather, below 50 K, it decreased as the temperature decreased.
This is indeed the susceptibility one would expect for AF correlated kagome spins, with similar
behavior inferred from $^{17}$O NMR data as well (Fig.~8, right) \cite{olariu}.  The latter measurements saw two different oxygen sites, which are now known to be
due to the influence of the copper defects sitting
on the zinc sites \cite{imai2,imai3}.  Subsequent ESR measurements indicated the presence of a substantial out-of-plane DM term \cite{zorko}.  With the
advent of single crystals, the susceptibility was studied in greater detail \cite{han3}.  The spin anisotropy changes sign from high temperatures to low
temperatures.  The former is thought to be due to an anisotropic exchange term, the latter due to the spin defects.  Application of a field of 2 Tesla,
though, can freeze the spins as seen also from $^{17}$O NMR \cite{jeong}.  The small energy scale related with this field scale is comparable to that
associated with the spin defects, indicating their field polarization.  Application of a pressure of 2.7 GPa leads to spin ordering \cite{kozlenko}.  A
$\sqrt{3} \times \sqrt{3}$ magnetic structure was inferred based on the powder samples, but this did not take into account any potential component along 00L (see below).

\begin{figure}
\includegraphics[width=0.49\hsize]{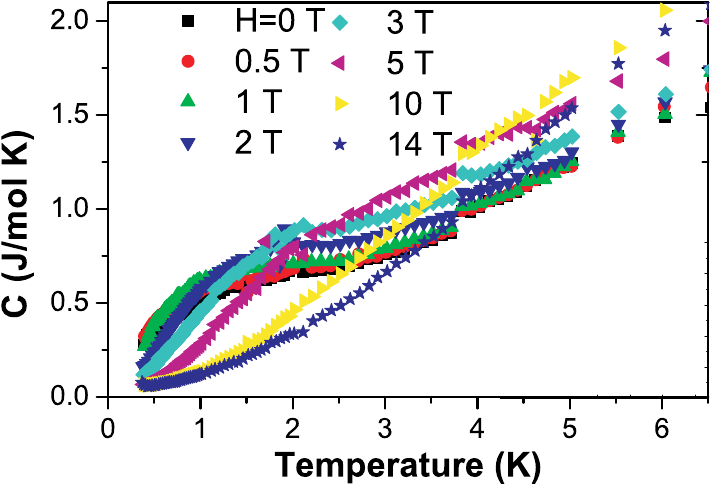}
\includegraphics[width=0.49\hsize]{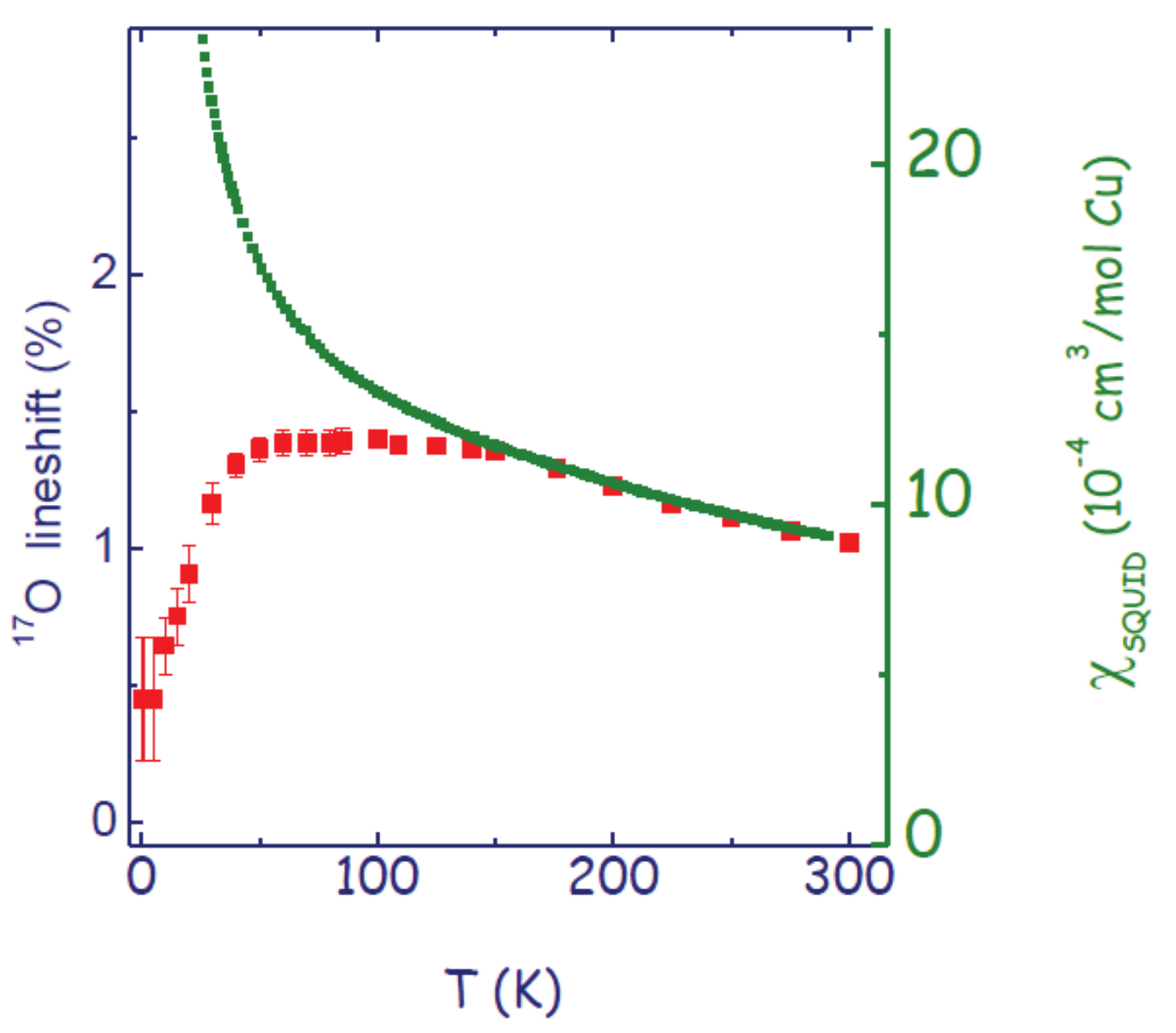}
\caption{(Color online) Left: Specific heat versus temperature for various magnetic fields \cite{helton1}.  The strong variation with field indicates
that the low temperature specific heat is primarily due to spin defects.
Right: Comparison of the bulk susceptibility (green curve) to that inferred from the $^{17}$O NMR line shift (red squares) for
herbertsmithite \cite{mendels2}.  At low temperatures, the former is dominated by the copper intersite defect spins, the latter by
the copper spins sitting in the kagome planes. 
}
\label{fig8}
\end{figure}

\subsection{Neutron scattering}

As alluded to above, neutron scattering is a powerful probe of the spin dynamics.  Elastic measurements found clearly defined Bragg peaks for $x$=0, 0.2 
and 0.4, with evidence for spin freezing below 5 K for $x$=0.66 (Fig.~9) \cite{lee}.  The inelastic measurements found that the `zero' mode at about 1.3 meV
for clinoatacamite is rapidly suppressed as a function of $x$ \cite{lee}.  For $x$=1, only a broad continuum in both energy and momentum was seen.  Power
law correlations as a function of energy were seen at 35 mK \cite{helton1} which subsequent measurements found to exhibit quantum critical scaling,
with the imaginary part of $\chi$ going as $\omega^{-\alpha} \tanh (\omega/\beta T)$ with $\alpha$=2/3 and $\beta$=5/3 (Fig.~10, left) \cite{helton2}, though a
scale-free behavior was claimed earlier \cite{devries2}.  Similar power laws were inferred by NMR \cite{olariu,imai1}, and were suggestive of the power
law correlations expected for a U(1) Dirac spin liquid \cite{ran}.
Detailed INS studies at very low energies, though, were consistent with defect behavior in this energy range, including their Zeeman shift in an
applied field \cite{nilsen}.

\begin{figure}
\includegraphics[width=0.7\hsize]{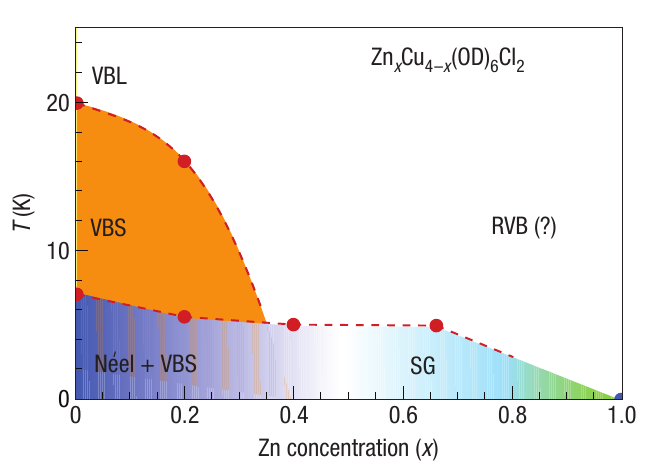}
\caption{(Color online)  Proposed phase diagram for Zn-paratacamite \cite{lee}.  `Neel' is the long-range canted AF order, `VBS' a speculated valence bond solid phase,
and `VBL' its melted version.  `SG' indicates spin-glass behavior, and `RVB' denotes the spin liquid phase.  The monoclinic to rhombohedral phase
transition occurs near $x$=1/3.  
}
\label{fig9}
\end{figure}

\begin{figure}
\includegraphics[width=0.49\hsize]{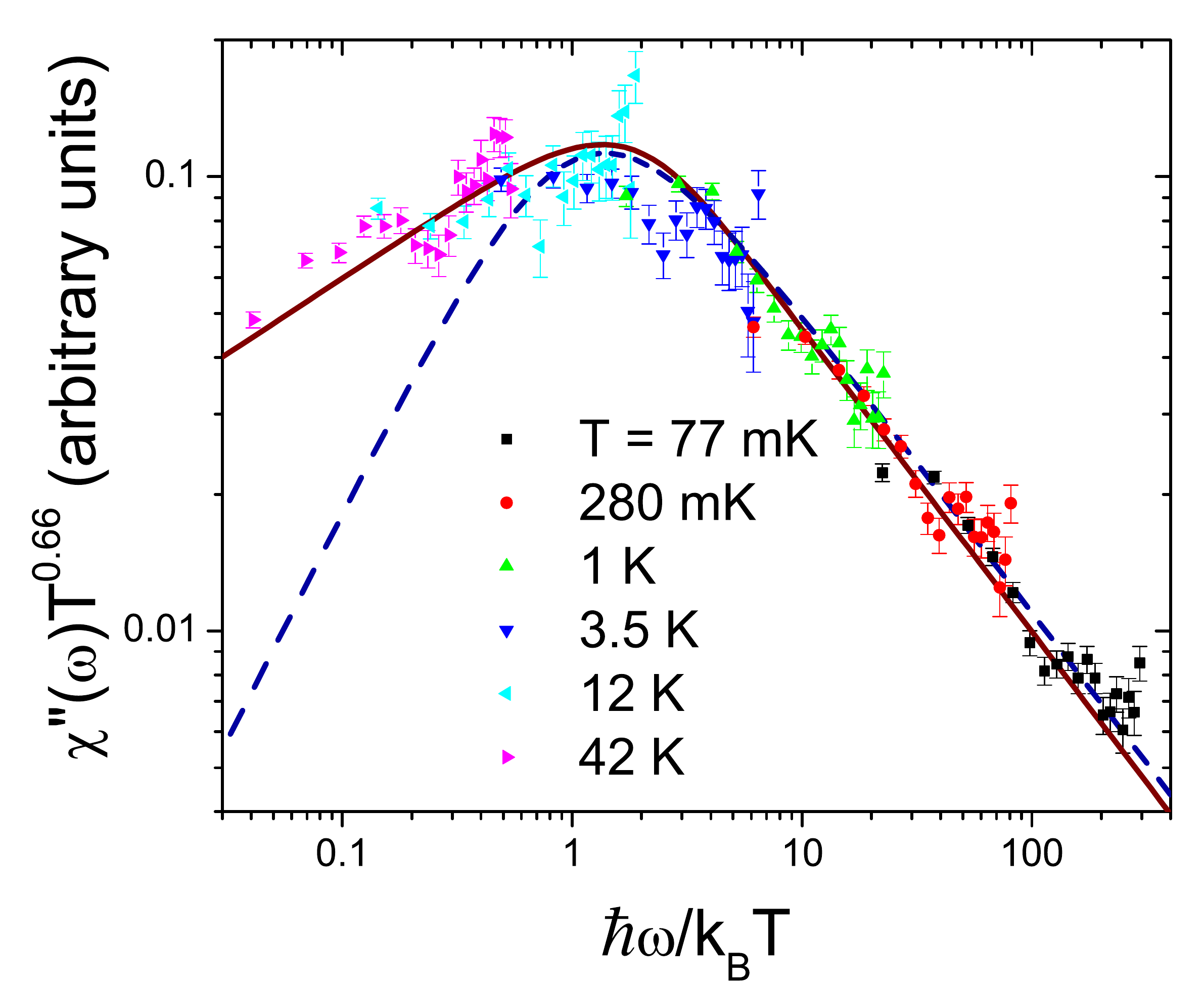}
\includegraphics[width=0.49\hsize]{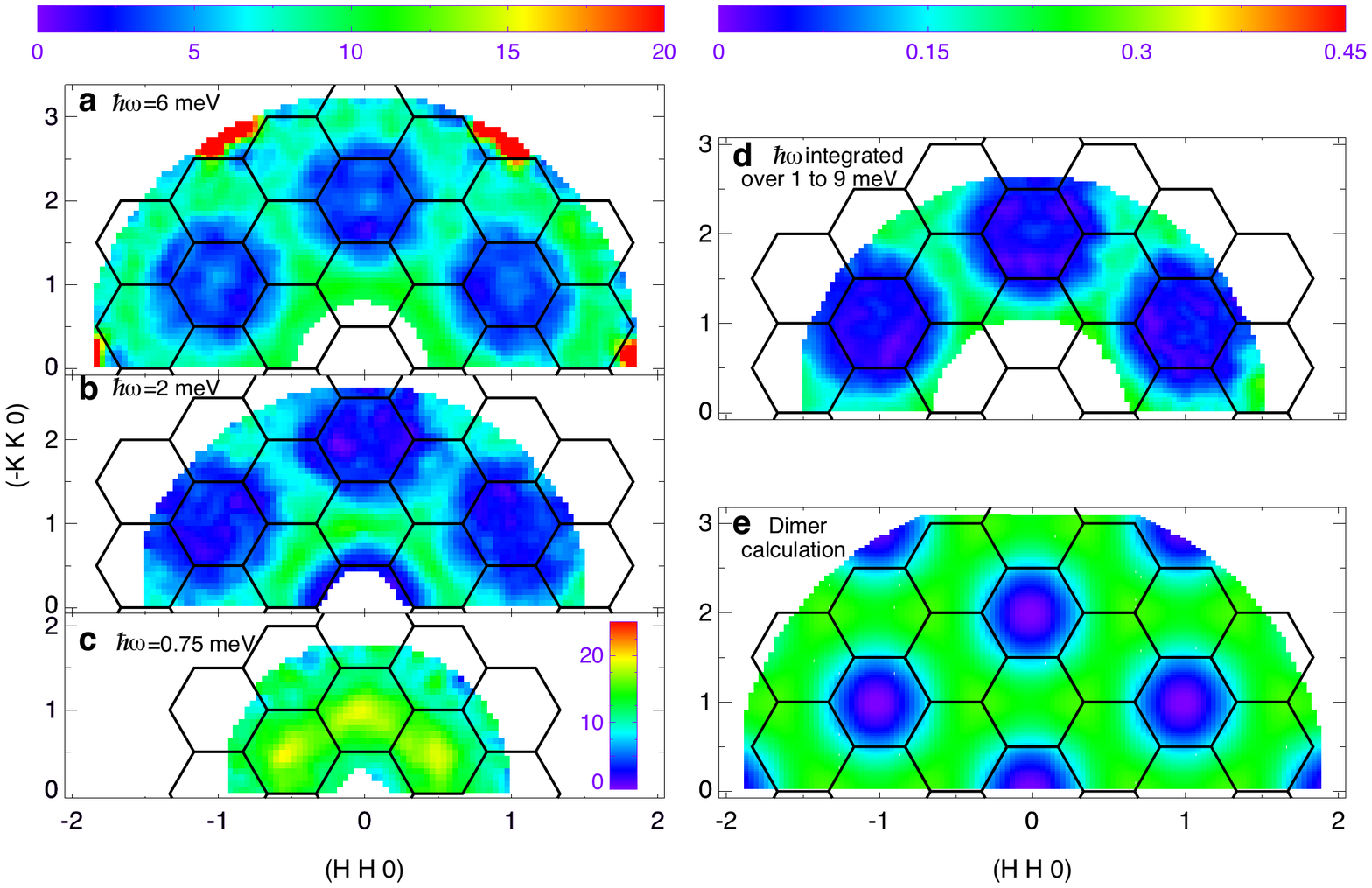}
\caption{(Color online)  Left: The imaginary part of the momentum-integrated dynamic spin susceptibility of herbertsmithite for various temperatures, demonstrating quantum critical-like scaling \cite{helton2}.  The solid curve is the scaling function described in the text.
Right: Momentum structure of the INS data for single crystal samples at 1.6 K for three different energies \cite{han1}.  Note the evolution from a near-neighbor dimer
pattern to a more spot-like pattern as the energy is reduced below 2 meV.
}
\label{fig10}
\end{figure}

\begin{figure}
\includegraphics[width=0.9\hsize]{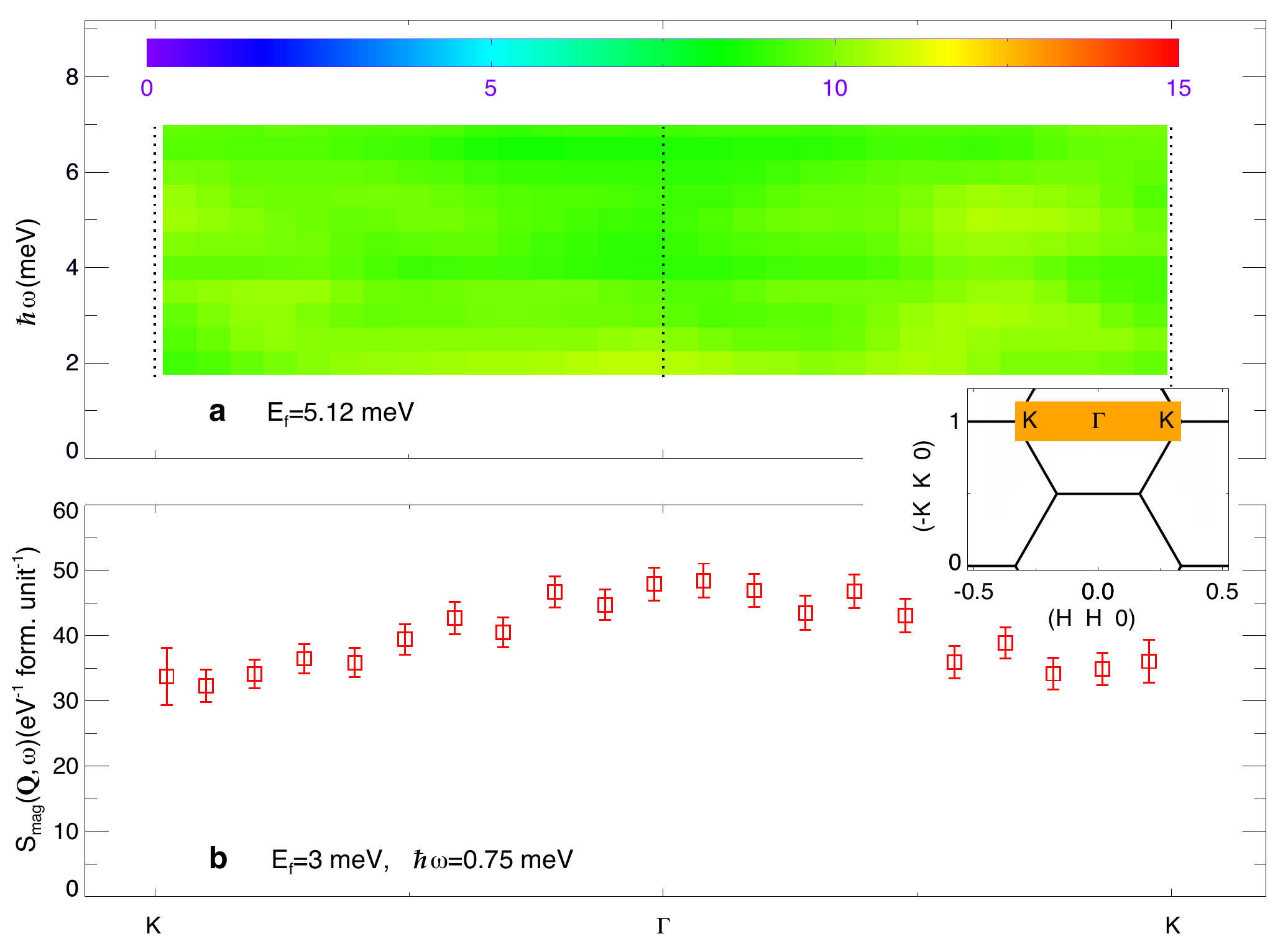}
\caption{(Color online)  INS spectra of herbertsmithite at 1.6 K along the $K-\Gamma-K$ direction as a function of energy \cite{han1}.
Note that the intensity is almost independent of energy and
momentum, in sharp contrast to the magnon-like excitations and `zero' modes seen in clinoatacamite.  This is evidence for a spinon continuum.
}
\label{fig11}
\end{figure}

The advent of single crystals has led to much richer results \cite{han1}.  For the most part, the spectra have a modest dependence on both momentum and
energy, in sharp contrast to the magnon-like excitations and `zero' modes seen in clinoatacamite (Fig.~11).  This has led to the feeling that this represents a true
spinon continuum (noting, though, that all theoretical models that have such a continuum exhibit a much stronger momentum/energy dependence than what is
observed \cite{punk}).  Above 1 meV or so, the momentum pattern is what one would expect for near-neighbor AF correlations within the kagome plane (Fig.~10, right),
with a correlation length of order 3 \AA,
though detailed fits indicate some contribution from longer-range exchange.  This is certainly consistent with ab initio calculations of the exchange
integrals \cite{jeschke}, which indicate a large near neighbor AF exchange (182 K), but a far weaker next-near neighbor AF exchange (3 K).  But below 1 meV, the pattern becomes 
more spot-like, with maxima at the center of the Brillouin zone (Fig.~10, right).  We focus on this point in the next Section.

\section{The Physics of Defects}

\subsection{Inelastic neutron scattering and NMR}

Naively, one might expect the zone center INS maxima below 1 meV \cite{han1} to simply be a reflection of a $q$=0 magnetic state, much like what is seen
in the iron jarosites \cite{grohol}.  That this is not the case is shown by new INS data taken in the (HHL) scattering plane (Fig.~12, bottom row) \cite{han4}.  Here, one
clearly sees a diffuse peak at (0,0,$\frac{3}{2}$).  Such (00L) peaks are inconsistent with the spins summing to zero on a kagome triangle (in the
iron jarosites, these peaks occur along (11L) instead \cite{matan-thesis}).  Another possibility would be ferromagnetic planes coupled antiferromagnetically
along the $c$-axis, but this is highly inconsistent with the large in-plane AF exchange interaction identified from various measurements.

\begin{figure}
\includegraphics[width=\hsize]{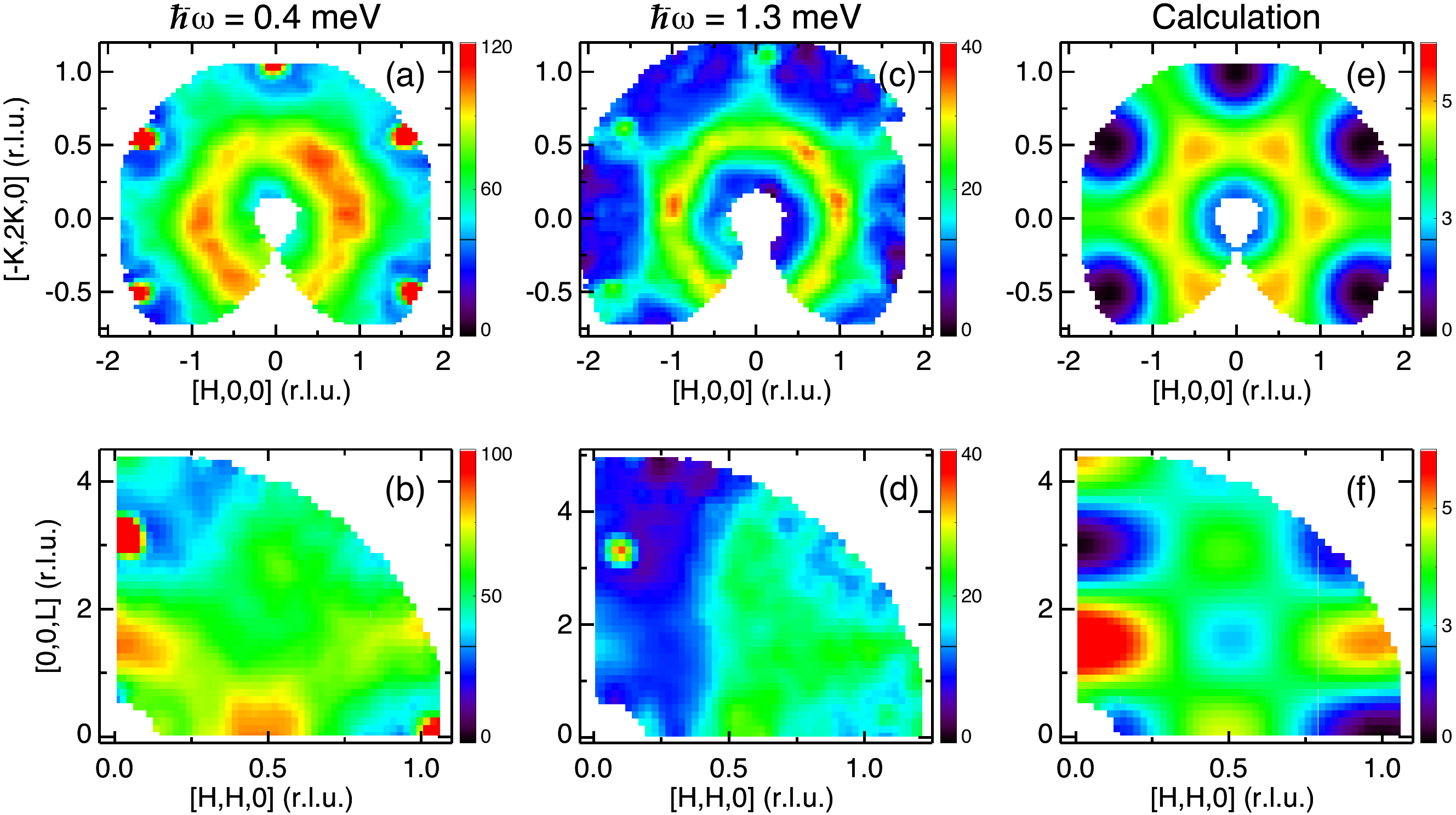}
\caption{(Color online) Momentum structure of the INS data at 0.4 meV and 1.3 meV for single crystal herbertsmithite at 2 K in the (HK0) scattering plane (top row)
and (HHL) scattering plane (bottom row) \cite{han4}.  The plots in the right column are the calculated structure factor for
near neighbor AF correlations between copper defects on the zinc sites, taking into account the copper form factor.
These correspond to correlations between the brown and the gray sites of Fig.~1, which sit in successive triangular planes.
}
\label{fig12}
\end{figure}

One is then forced to conclude that this pattern has something to do with the defects.  Indeed, AF correlations between near neighbor defect sites (which sit in
neighboring triangular planes) give rise to such a pattern (Fig.~12, right column) \cite{han4}.  Such correlations can be motivated by the known magnetic structure
of clinoatacamite.  This implies that the copper defects on the zinc sites locally distort the surrounding matrix (due to the Jahn-Teller effect).  This effect has been
inferred as well from $^{35}$Cl NMR data \cite{imai2,imai3} and could be further investigated by such techniques as PDF (pair distribution function) or
EXAFS.  Exploiting the differing momentum dependences of the kagome
and defect spins, one can estimate that a spin gap of about 0.7 meV exists for the kagome spins.  The same value is found when modeling momentum
integrated data as the sum of a damped harmonic oscillator (previously used to model the defect spins in an earlier INS study \cite{nilsen}) and a gapped
kagome contribution (Fig.~13) \cite{han4}.

\begin{figure}
\includegraphics[width=0.6\hsize]{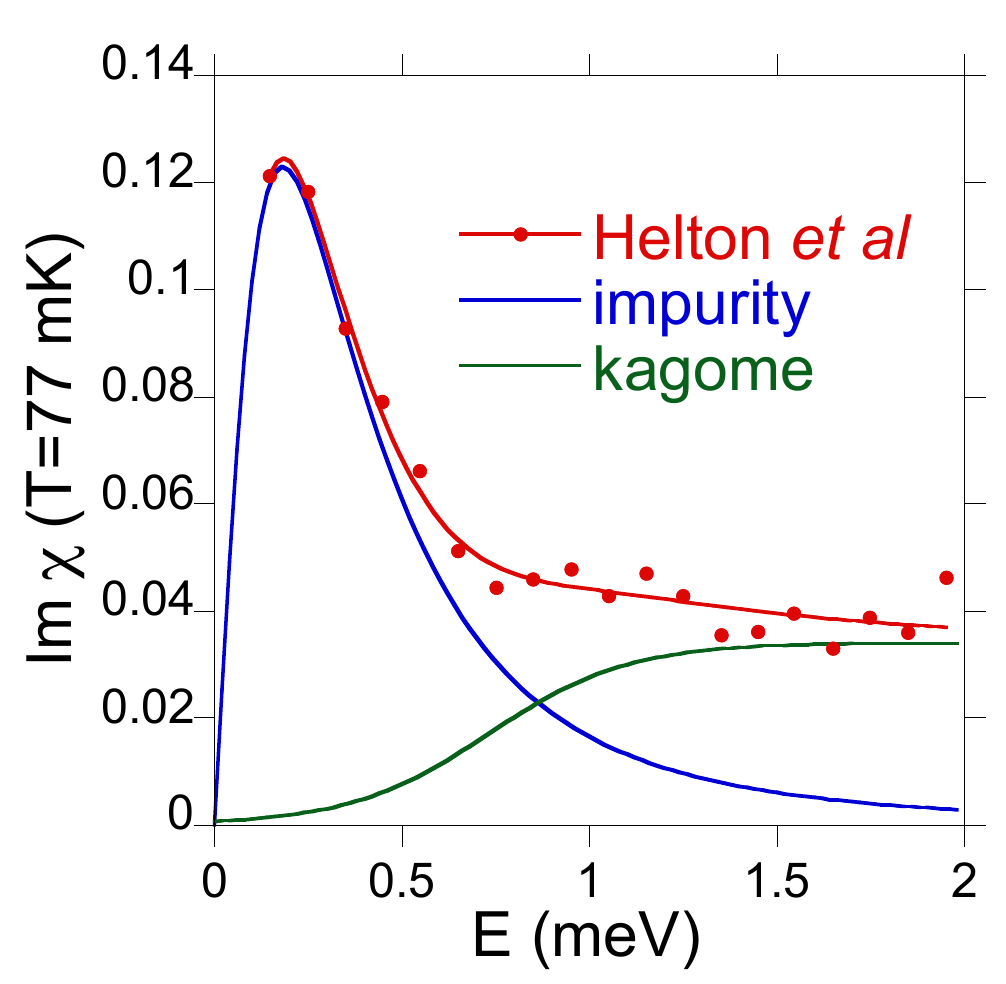}
\caption{(Color online) Momentum-integrated INS data \cite{helton2} for herbertsmithite fit to a sum of a low energy damped harmonic oscillator representing the defect
spins \cite{nilsen} and a gapped continuum representing the kagome spins \cite{han4}.  The spin gap from the latter is 0.73 meV, close to that inferred
from $^{17}$O NMR data \cite{imai3}.
}
\label{fig13}
\end{figure}

About the same value for the spin gap was earlier estimated from NMR data \cite{imai3}.  This study found three different oxygen sites, with site occupations based
on earlier x-ray studies \cite{freedman} found to be consistent with the number of near neighbor and next-near neighbor oxygens about a given copper defect on
the zinc sites, with the majority of the oxygen sites (roughly 59\%) being largely unaffected by the defects (the same numbers were inferred from earlier NMR studies as
well \cite{olariu,imai2}).  From this last type, one can estimate a kagome spin gap of about 0.05$J$, roughly consistent with DMRG and exact diagonalization studies of the Heisenberg kagome lattice, and consistent with the later INS studies mentioned above.

\subsection{Quantum criticality versus random bonds}

These findings touch on a long standing debate in physics concerning the quantum critical behavior observed by neutrons in several strongly correlated electron
systems.  The same scaling function used to fit heavy fermion 5f materials \cite{aronson}
was also used by \textcite{helton2} to fit their INS data on herbertsmithite (Fig.~10, left).  In the heavy
fermion field, it was advocated that defects could cause this scaling by inducing a random distribution of Kondo temperatures \cite{neto}, and in fact the scaling form used in \textcite{helton2} has been connected to
that of the random Heisenberg model by \textcite{singh}.  Detailed studies of the random Heisenberg model on a kagome lattice have claimed to be
consistent with the herbertsmithite INS data, in particular an energy independent and relatively momentum independent continuum (with a low energy intensity
upturn) \cite{kawamura},
but such studies do not take into account the intersite defect nature of the actual low energy data.  What is clear from the analysis of \textcite{han4}
is that the kagome spins appear to be remarkably unaffected by the defect spins, likely due to the fact that the probability that one has a defect spin on
both sides of a kagome spin is only about 2\%.  This indicates that the kagome contribution to the INS data is likely a pristine representation of an ideal
kagome lattice, making the case for a spinon continuum a reasonable one.

Although the momentum dependent correlations among the defects is fascinating (who would have expected 3D correlations for such a quasi-2D material?),
the fact remains that it would be nice to find a material analogue where the defect concentration was not so high.  Unfortunately, data on the magnesium variant
of herbertsmithite (tondiite) appear to be plagued by the same defect problems as its zinc sibling \cite{kermarrec}, somewhat surprising given its smaller ionic radius.
As we mentioned earlier, the introduction of cadmium makes things even worse because of its larger ionic radius \cite{mcqueen}.  On the other hand, it is
possible other 2+ ions would ameliorate these effects, so this is definitely worth exploring.

\section{The Future}

For experiment, there are several potential directions to pursue.  The first is to bring more techniques to bear.  Many of the probes used for the cuprates
have yet to be performed for herbertsmithite - the obvious examples are angle resolved photoemission, x-ray absorption, scanning tunneling microscopy, infrared
conductivity, and both electric and thermal transport.  As a consequence, the actual electronic structure of herbertsmithite is not known.  Although one might expect many
similarities to the cuprates, the fact that these materials are hydroxychlorides instead of oxides means there will be many differences as well.
So far, Raman has indicated a spin background somewhat reminiscent of cuprates \cite{wulferding}, and the in-plane THz conductivity sees
field-independent power-law behavior \cite{gedik} as expected for a gapless (or near gapless) spin liquid \cite{potter}.  Still, we have
a long way to go before we have as thorough an understanding for herbertsmithite as we do for stoichiometric cuprates.

The second is the investigation of related materials.  For instance, there are a large number of compounds, particularly minerals, which have only
been studied from a crystallographic point of view.  As an example, the copper tellurium oxide quetzalcoatlite,
Zn$_6$Cu$_3$(TeO$_3$)$_2$O$_6$(OH)$_6$(Ag$_x$Pb$_y$)Cl$_{x+2y}$, is composed of perfect copper kagome layers exhibiting AA stacking \cite{quetzal}.
Unfortunately, the natural crystals are only micron size (no synthetic studies exist).  The bond pathway in the kagome layers is of the type Cu-O-Te-O-Cu,
and therefore exhibits
super-superexchange (Cu-O-O-Cu), which is weaker than superexchange (Cu-O-Cu), though a number of copper tellurium oxides are known which have sizable magnetic
transition temperatures.  The reason this particular mineral is brought up is that, like herbertsmithite, the layers are connected by zinc ions, but in quetzalcoatlite,
the zinc is tetrahedrally coordinated instead, meaning the issue of copper on the zinc sites that plagues herbertsmithite should not be an issue for this material.

\subsection{Doping herbertsmithite}

The real frontier, though, is chemical doping.  \textcite{mazin} have shown that the band structure of this material should have a Dirac point at 100\%
electron doping that could be achieved by substitution of zinc by gallium, which has a comparable ionic radius.  They predict that such a material
will also have f-wave superconductivity because of the triangular nature of the lattice.  For the hole-doped case (say, by replacing zinc by lithium),
one expects topological flat bands to come into play \cite{guterding}.  The latter study also extensively checked the defect energetics, which suggests
that a number of 1+ or 3+ ions could be substituted for zinc.

The reality so far, though, has been disappointing.  Attempts have been made to `dope' herbertsmithite by cation substitution, electrochemically,
and even by irradiation \cite{bartlett-thesis,nytko-thesis}.  The net result is that either nothing happens, or the material decomposes, typically
leading to the formation of CuO (with the zinc component sometimes coming out as simonkolleite).  This is not difficult to understand.  Attempts
to substitute a different ionicity on the zinc site should lead to (OH)$^-$ either becoming H$_2$O or O$^{--}$, causing the material
to fall apart.  This is connected to the low formation temperature of these materials, and the existence of multiple polymorphs.
Even for non-doped materials, what forms is very sensitive to the ratio of the various ions in solution, which determines the coordination shell
around the copper ions (Fig.~14) \cite{sharkey,tecton}.  Still, this subject has been given far less attention than it should.  A doped version of herbertsmithite,
particularly a metallic variant, would be a significant discovery.
In that context, recently, lithium has been intercalated into herbertsmithite \cite{kelly}, but the material remains insulating, perhaps due to
localization of the doped carriers.

\begin{figure}
\includegraphics[width=\hsize]{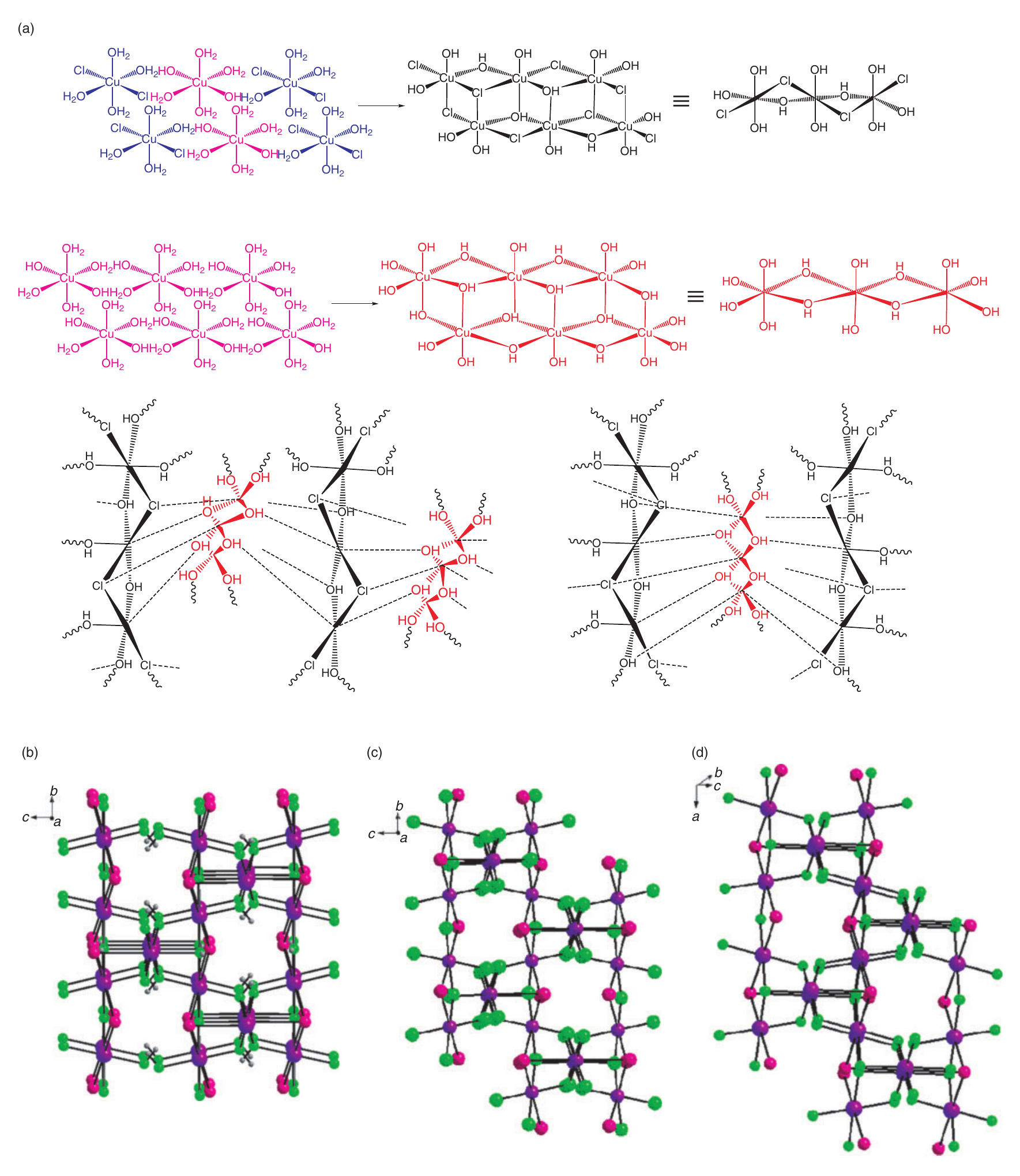}
\caption{(Color online) (a) Synthesis of different copper minerals determined by the relative concentration of various copper tectons in solution \cite{tecton}.
The different polymorphs of Cu$_4$(OH)$_6$Cl$_2$ are formed from chains resulting from the condensation of
tecton I, Cu(H$_2$O)$_4$Cl$_2$, and tecton II, Cu(H$_2$O)$_4$(OH)$_2$,
in the ratio of 1 to 3, which then assemble in different ways leading to the formation of (b) atacamite, (c) paratacamite and (d) clinoatacamite.
}
\label{fig14}
\end{figure}

\subsection{Topological degeneracy and fractionalized excitations}

A final frontier, though, is theory.  So far, exact diagonalization studies have been limited to 48 sites, with most of the reported results
for 36 sites or less.  Although there are extensive DMRG results, studies of models other than the ideal near neighbor Heisenberg model on
the kagome lattice are still in their infancy, even before mentioning PEPS, MERA or quantum Monte Carlo.  Various theory papers have pointed to
the importance of longer range exchange, DM and anisotropic exchange, and interlayer interactions \cite{cepas,jeschke}.  And the incorporation
of defects into theoretical models has only seen limited attention.

Perhaps the most profound discovery would be a proof of either topological degeneracy, or fractionalized excitations.  For the fractional quantum Hall
effect, this took a long time despite the very controlled nature of GaAs heterostructures.  Ultimately, if such could be done in herbertsmithite,
it will give us a much better understanding of what it means to be a quantum spin liquid.

\begin{acknowledgments}
This work was supported by the Materials Sciences and Engineering Division, Basic Energy Sciences, Office of Science, US DOE.
The author would like to thank one of his collaborators, Tian-Heng Han, whose interest in herbertsmithite helped to inspire my own.
\end{acknowledgments}

\end{document}